\documentclass[prc,amsmath,twocolumn,superscriptaddress,preprintnumbers]{revtex4}

\usepackage{gensymb}
\usepackage{graphicx,color}
\usepackage{amssymb}
\usepackage{enumerate}
\usepackage{verbatim}
\usepackage{natbib}
\usepackage{amsmath}

\begin{document}
  \newcommand {\nc} {\newcommand}
  \nc {\Sec} [1] {Sec.~\ref{#1}}
  \nc {\IR} [1] {\textcolor{red}{#1}} 
  \nc {\IB} [1] {\textcolor{blue}{#1}}
  \nc {\cgmf}{$\mathtt{CGMF}$}
  \nc {\coh}{$\mathtt{CoH}$}
  \nc {\beoh}{$\mathtt{BeoH}$}
  \nc {\hfd}{$\mathtt{HF^3D}$}

\title{Extension of the Hauser-Feshbach Fission Fragment Decay Model to Multi-Chance Fission}

\author{A.~E.~Lovell}
\email{lovell@lanl.gov}
\affiliation{Los Alamos National Laboratory, Los Alamos, NM 87545, USA}
\author{T.~Kawano}
\email{kawano@lanl.gov}
\affiliation{Los Alamos National Laboratory, Los Alamos, NM 87545, USA}
\author{S.~Okumura}
\affiliation{NAPC--Nuclear Data Section, International Atomic Energy Agency, Vienna A-1400, Austria}
\author{I.~Stetcu}
\affiliation{Los Alamos National Laboratory, Los Alamos, NM 87545, USA}
\author{M.~R.~Mumpower}
\affiliation{Los Alamos National Laboratory, Los Alamos, NM 87545, USA}
\author{P.~Talou}
\affiliation{Los Alamos National Laboratory, Los Alamos, NM 87545, USA}

\date{\today}


\begin{abstract}
The Hauser-Feshbach fission fragment decay model, \hfd, which calculates the statistical decay of fission fragments, has been expanded to include multi-chance fission, up to neutron incident energies of 20 MeV.  The deterministic decay takes as input pre-scission quantities--fission probabilities and the average energy causing fission--and post-scission quantities--yields in mass, charge, total kinetic energy, spin, and parity.  From these fission fragment initial conditions, the full decay is followed through both prompt and delayed particle emissions, allowing for the calculation of prompt neutron and $\gamma$ properties, such as multiplicity and energy distributions, both independent and cumulative fission yields, and delayed neutron observables.  In this work, we describe the implementation of multi-chance fission into the \hfd{} model, and show an example of prompt and delayed quantities beyond first-chance fission, using the example of neutron-induced fission on $^{235}$U.  
This expansion represents significant progress in consistently modeling the emission of prompt and delayed particles from fissile systems.  
\end{abstract}

\preprint{LA-UR-20-27423}

\maketitle
\section{Introduction}
\label{sec:intro}


In the over 75 years since fission was discovered, there are still many unknowns about the process, and the modeling is complicated, particularly to study the process from the formation of the compound nucleus through the prompt and delayed emission of neutrons and $\gamma$ rays.  Typically, this modeling is broken into two or three components, modeling the descent of the compound nucleus from saddle to scission (microscopic or macroscopic-microscopic models), the emission of prompt neutrons and $\gamma$ rays (deterministic or Monte Carlo), and the delayed particle emission.  For models describing the prompt and delayed decay, input on the initial conditions of the fission fragments is needed, typically yields in mass, charge, kinetic energy, spin, and parity, $Y(A,Z,\mathrm{TKE},J,\pi)$, where $A$, $Z$, $J$, and $\pi$ are the fission fragment mass, atomic number, spin, and parity respectively, and TKE is the total kinetic energy of both fission fragments.  Part of this initial distribution can be calculated through the microscopic or macroscopic-microscopic models, e.g. Refs.~\cite{Schunck2014,Sierk2017,Verriere2019,Regnier2016,Moller2015,Usang2019,Ishizuka2017,Mumpower2020}, although currently, the quality of these models is not high enough to serve as input for the prompt and delayed decay, at least for well-known fissioning systems.  Instead, it is standard practice for $Y(A,Z,\mathrm{TKE},J,\pi)$ to be parametrized and fit to experimental data such as in Refs.~\cite{Wahl2002,Brosa1990,GEF,Becker2013,FREYA1,FREYA2,FIFRELIN,Okumura2018}.  

Many models used for the calculation and evaluation of quantities relevant to applications, such as reactor design, understanding detector response in nuclear physics measurements~\cite{Kelly2019,Jandel2015}, or stockpile stewardship, are independent of one another so there is no connection between the fission product yields (FPYs), multiplicity distributions, or energy spectra.  This independence has been shown to cause discrepancies between evaluated quantities~\cite{Jaffke2018}.  Thus, a great investment has been made into developing models that can calculate these quantities consistently.  These models are either Monte Carlo, such as \cgmf{}~\cite{Becker2013,CGMF}, $\mathtt{FREYA}$~\cite{FREYA1,FREYA2}, $\mathtt{GEF}$~\cite{GEF} and $\mathtt{FIFRELIN}$~\cite{FIFRELIN} or deterministic, as is \beoh{}~\cite{Okumura2018}, which is the focus of our current work.  The Monte Carlo models allow for event-by-event correlations between fission fragments and emitted particles, however, they can be time-consuming, especially to fully sample the lowest yields of the initial yield distribution.  The deterministic models can calculate these low-yield observables with the same computational accuracy as for high-yield regions but at the cost of losing track of the correlation between the observables.

The deterministic \hfd{} (Hauser-Feshbach Fission Fragment Decay) model implemented in \beoh{} was studied by Okumura, et al. \cite{Okumura2018} for neutron-induced fission of $^{235}$U below the threshold for second-chance fission.  They found that by fitting mass and kinetic energy distributions to experimentally measured data and optimizing the excitation energy splitting along with the spin cut-off parameter, they were able to reproduce many experimentally measured prompt observables, such as independent yields, TKE distribution, $\overline{\nu}_p$ (average number of prompt neutrons per fission) as a function of incident energy and mass, the neutron multiplicity distribution, and the isomeric ratios. The energy dependence of these quantities can be attributed to the dynamical process of the compound nucleus toward the scission point:  (a) the most probable fission path after the second barrier spreads along the most probable path, hence the asymmetric components in $Y(A,Z)$ will have wider width, and the peaks of distributions will be lower due to normalization, (b) the energy sharing mechanism, which determines the initial excitation energies in both fission fragments, is affected by the compound nucleus excitation energy, and (c) the even-odd effect~\cite{Wahl2002} in the charge distribution $P(Z|A)$ to obtain $Y(A,Z)$ might decrease when the fissioning system has more excitation energy. 

In this work, we extend the \hfd{} model beyond the second- and third-chance fission thresholds, where we expect more notable signatures of the energy-dependence of fission observable, such as TKE, independent and cumulative FPYs, $\overline{\nu}_p$ and $\overline{\nu}_d$ (average number of prompt and delayed neutrons), and so forth. In fact, experimental TKE reveals the multi-chance effect~\cite{Duke2015,ANLNDM64,Zoller1995}, and Lestone and Strother~\cite{Lestone2014} interpret the structure seen in the experimental TKE by the multi-chance fission model. We provide the extended \hfd{} model including multi-chance fission, and calculate fission observables up to 20 MeV incident energy. 


\section{Theory}
\label{sec:theory}



When the excitation energy of the fissioning compound nucleus is higher than one neutron separation energy $S_n$, the prompt neutrons and $\gamma$ rays are produced by both the compound nucleus before scission and the two fission fragments. The pre-fission neutrons are weakly coupled with the post scission phenomena by the energy conservation, as $S_n$ and the pre-fission neutron kinetic energy removes the compound nucleus excitation energy to some extent. Here we model these two stages (before and after scission) consistently in the Hauser-Feshbach framework~\cite{HauserFeshbach}. The compound nucleus decay process can be followed by either the deterministic method or Monte Carlo, but the formalism stays the same.  

\subsection{Pre-scission calculations}

The pre-scission calculations that are needed to estimate the multi-chance fission probabilities and excitation energy causing fission are performed with the statistical Hauser-Feshbach code, \coh{}~\cite{CoH}.  \coh{} calculates the fission cross sections $\sigma_f(m)$ for the $m^\mathrm{th}$-chance ($m$=1, 2, ...) by the statistical Hauser-Feshbach theory, where the maximum $m$ depends on the fission barriers and excitation energy available in the compound system. The multi-chance fission probability $P_f(m)$ is calculated from a ratio of the multi-chance fission cross sections, $\sigma_f(m)$,
\begin{equation}
P_f (m) = \frac{\sigma_f (m)}{\sigma_f} \ ,
\label{eqn:MCFprob}
\end{equation}
where the total fission cross section $\sigma_f$ is the sum of the cross section of the $m^\mathrm{th}$-chance fission channels, $\sigma_f = \sum \limits_m \sigma_f(m)$. 

Furthermore, a partial $\sigma_f(m)$ is defined as a function of the compound nucleus excitation energy $E_x$, and the average excitation energy $\langle E_f\rangle(m)$ in the $m^\mathrm{th}$ compound nucleus, where $m^\mathrm{th}$-chance fission is most probable, is
given by
\begin{equation}
\langle E_f \rangle (m) = \frac{\displaystyle\int \sigma_f(m,E_x) E_x dE_x}
                               {\displaystyle\int \sigma_f(m,E_x) dE_x} \ ,
\end{equation}
where the integration range over excitation energy, $E_x$, is given by energy conservation, and $\sigma_f(m,E_x)$ is the partial fission cross section from the compound state at $E_x$. The partial fission cross section is defined as the product of a formation cross section of the compound state $\sigma_P(m,E_x)$ and a branching ratio to the fission channel, which is schematically written in the statistical model in a simplified way by
\begin{equation}
  \sigma_f(m,E_x) = \sigma_P(m,E_x) 
                    \frac{T_f(m,E_x)}
                         {T_n + T_\gamma + T_f(m,E_x)} \ .
  \label{eqn:SigF}
\end{equation}
$T_n$, $T_\gamma$, and $T_f$ are the lumped transmission coefficients for neutron emission, $\gamma$ emission, and fission, respectively.  The level density and the discrete states in a residual nucleus are implicitly involved in calculating the $T$'s, thus the calculation is performed by considering the spin and parity conservation at each decay stage.  We assume that the only decay processes allowed are emission by neutrons and $\gamma$ rays or fission, which is justified by the energy range considered in this work.

In general, the probability of fission increases as the compound nucleus excitation energy rises. An incident neutron with the center-of-mass energy of $E_\mathrm{inc} ^\prime$ forms a compound nucleus at $E_{\rm  max}(m=1) = S_n(1) + E_\mathrm{inc} ^\prime$, which is the highest available energy of the initial compound nucleus (first-chance fission), and the average excitation energy causing fission $\langle E_f\rangle(1)$ coincides with $E_{\rm max}(1)$.  Beyond the second-chance threshold, $m \ge 2$, fission may occur at lower excitation energies than the allowed maximum energy of
\begin{eqnarray}
  E_{\rm max}(m)
  &=& E_{\rm max}(1) - \sum_m S_n(m-1) \quad (m \ge 2) \\
  &\ge& \langle E_f\rangle(m) \ ,
\end{eqnarray}
due to the distribution of populated nuclear states after neutron evaporation, as well as a small probability of emitting $\gamma$ rays before pre-fission neutron emission. It is important to note that even though fission takes place by interacting with a high-energy neutron, the average excitation energy causing fission along the multi-chance process, $\sum_m P_m \langle E_f\rangle(m)$, reminds relatively low, e.g. 10~MeV or so, as the pre-fission neutrons take away energy from the compound nucleus, cooling the system. This feature will be discussed further in Section \ref{sec:results}.


\subsection{Post-scission calculations}

\beoh{} contains the Hauser-Feshbach Fission Fragment Decay (\hfd{}) model used to calculate the emission of prompt particles.  In this model, each fission fragment is treated as a compound nucleus and its decay through emission of neutrons and $\gamma$ rays is calculated using the statistical Hauser-Feshbach theory~\cite{HauserFeshbach}.  To calculate the statistical decay, the initial conditions of each fission fragment are needed, primarily the yield in mass, charge, kinetic energy, spin, and parity, $Y(A,Z,\mathrm{TKE},J,\pi)$.  This yield distribution is built up of partial yield distributions, namely $Y(A)$,  $Y(Z|A)$, $Y(\mathrm{TKE}|A)$, and $Y(J,\pi)$. 

The fission fragment mass yield, $Y(A)$, is incident-energy dependent and constructed as the sum of three Gaussians,
\begin{equation}
Y(A|E_\mathrm{inc}) = G_0 (A|E_\mathrm{inc}) + G_1 (A|E_\mathrm{inc}) + G_2 (A|E_\mathrm{inc}),
\end{equation}
with
\begin{equation}
G_0(A|E_\mathrm{inc})
   = \frac{W_0(E_\mathrm{inc})}
          {\sqrt{2\pi} \sigma_0(E_\mathrm{inc})}
     \exp \left [ \frac{-(A-A_c/2)^2}{2\sigma_0 (E_\mathrm{inc})^2} \right ],
\end{equation}
and
\begin{eqnarray}
G_i (A|E_\mathrm{inc}) 
   &=& \frac{W_i(E_\mathrm{inc})}
            {\sqrt{2\pi} \sigma_i(E_\mathrm{inc})} \nonumber \\
   &\times& \left \{
               \exp \left[
                    \frac{-(A-\mu_i(E_\mathrm{inc}))^2}{2\sigma_i (E_\mathrm{inc})^2}
                    \right]
            \right. \nonumber \\
   &+& \left.
               \exp \left[
                    \frac{-(A-(A_c-\mu_i(E_\mathrm{inc})))^2}{2\sigma_i (E_\mathrm{inc})^2}
                    \right]
            \right \} \nonumber \\
   &~& (i=1,2) \ .
\end{eqnarray}
Each of the weights, means, and widths are allowed to be energy-dependent with
\begin{equation}
W_i (E_\mathrm{inc}) = \frac{1}{1+\exp{[(E_\mathrm{inc}-w_i^a)/w_i^b}]} \ ,
\end{equation}
\begin{equation}
\mu_i (E_\mathrm{inc}) = A_c/2 + \mu_i^a + \mu_i^b E_\mathrm{inc} \ ,
\end{equation}
\noindent and
\begin{equation}
\sigma_i (E_\mathrm{inc}) = \sigma_i^a + \sigma_i^b E_\mathrm{inc}  \quad (i=1,2) \ .
\end{equation}
In each of the above equations, $A_c$ is the mass number of the compound nucleus undergoing fission.  The weight of the symmetric mode, $W_0$, is determined by the normalization condition, $2 = W_0 + 2W_1 + 2W_2$, and the mean of the symmetric mode is fixed at the symmetric mass, $\mu_0 = A_c/2$.

The charge distribution, $Y(Z|A)$, is taken from the Wahl systematics~\cite{Wahl2002}.  In addition, we allow for a scaling of the $F_Z$ and $F_N$ parameters in the Wahl systematics to better reproduce the odd-even staggering of the cumulative fission yields.  

The kinetic energy as a function of mass is parametrized as
\begin{eqnarray}
\mathrm{TKE}(A_h)
   &=& (p_0 - p_1 A_h)\nonumber \\
   &\times& \left \{
                1 - p_2 \exp \left ( -\frac{(A_h - A_c/2)^2}{p_3} \right )
            \right \} \nonumber \\
   &+& \epsilon _\mathrm{TKE} \ ,
\label{eqn:TKEA}
\end{eqnarray}
the same as in~\cite{Okumura2018}, where $ [ p_0, p_1, p_2, p_3 ] $ are fitting parameters, and $A_h$ is the mass of the heavy fragment.  The small term $\epsilon _\mathrm{TKE}$ is included to ensure that the average of $\mathrm{TKE}(A_h)$ agrees with $\langle \mathrm{TKE} \rangle$ (to be discussed in Eq.~(\ref{eqn:TKEE})).  In this way, the energy dependence of $\mathrm{TKE}(A_h)$ is taken into account in an approximate way.  This overall shift in $\mathrm{TKE}(A_h)$ with incident neutron energy is consistent with experimental data, which mostly shows only a change in the magnitude and not a change in shape for $\mathrm{TKE}(A_h)$ as the incident neutron energy increases, e.g.~\cite{Dyachenko1968,Akimov1971}.  
In addition, the width of the TKE distribution is constructed as a function of $A_h$,
\begin{equation}
\sigma _\mathrm{TKE} (A_h) = s_0 - s_1 \exp \left [ -s_2(A_h - A_c/2)^2\right ] \ ,
\label{eqn:sigmaTKE}
\end{equation}
consistent with the shape of experimental data, as in Ref.~\cite{Baba1997}.

The TKE from Eq.~(\ref{eqn:TKEA}) is an average over all allowed charges for a single mass number.  To calculate the TKE for a given fragment pair, $(Z_l,A_l)$ and the corresponding $(Z_h,A_h)$, we assume that the TKE is proportional to the product of the charges
\begin{equation}
\mathrm{TKE}(Z_l,A_l) = \mathrm{TKE}(Z_h,A_h) = \mathrm{TKE}(A_h) \frac{Z_lZ_h}{D},
\end{equation}
where $D$ is a normalization constant, determined by summing over all possible $Z_lZ_h$ pairs.  This distribution of the TKE does not take into account any deformation in the fission fragments and is consistent with a description of the fragments having no kinetic energy at scission~\cite{Bulgac2019}.  The total excitation energy, TXE, for this fragment split is then calculated from energy conservation
\begin{equation}
\mathrm{TXE} = Q - \mathrm{TKE},
\end{equation}
where $Q$ is the $Q$-value of the fragment split.  When TXE is less than zero, these unphysical pairs are eliminated.  

The total kinetic energy is split between the two fission fragments based on kinematics,
\begin{equation}
\mathrm{KE}_l = \mathrm{TKE}(Z_l,A_l) \frac{A_h}{A_l+A_h},
\end{equation}
for the light fragment and
\begin{equation}
\mathrm{KE}_h = \mathrm{TKE}(Z_h,A_h) \frac{A_l}{A_l+A_h},
\end{equation}
for the heavy fragment.  The excitation energy is then distributed between the two fission fragments based on a ratio of temperatures, the anisothermal parameter $R_T$~\cite{Ohsawa1999,Ohsawa2000},
\begin{equation}
R_T = \frac{T_l}{T_h} = \sqrt{\frac{U_l}{U_h} \frac{a_h(U_h)}{a_l(U_l)}},
\label{eqn:TXEdist}
\end{equation}

\noindent where $a(U)$ is the energy-dependent level density parameter, calculated by the Gilbert-Cameron level density model~\cite{Gilbert1965}, and the excitation energy $U$ is corrected by the pairing energy ($U = E_x - \Delta$).  The excitation energies in the light and heavy fragments, $U_l$ and $U_h$, are iteratively searched over to satisfy Eq.~(\ref{eqn:TXEdist}).  To better reproduce $\overline{\nu}$ as a function of mass, $R_T$ is taken to be mass-dependent.

The excitation energy distribution, spin distribution, and parity distribution over each fission fragment need to be known to perform the statistical decay.  Because of the complexity of the fission process, we assume that even and odd parity states are equally produced after scission and take the parity distribution to be $1/2$.  A joint distribution over the excitation energy, $E_x$, and spin, $J$, are created, the initial population $P_{l,h}(E_x,J,\pi)$, normalized such that
\begin{eqnarray}
 &~& \sum_{J \pi} \int P_{l,h}(E_x,J,\pi) dE_x \nonumber \\
 &=& \sum_{J \pi} \frac{1}{2} \int G_{l,h}(E_x) R(J) dE_x \nonumber \\
 &=& 1 \ .
\end{eqnarray}
The excitation energy distribution, $G_{l,h}(E_x)$, is taken to be a
Gaussian,
\begin{equation}
  G_{l,h}(E_x) = \frac{1}{\sqrt{2\pi}\delta _{l,h}}
                \exp \left \{
                             - \frac{(E_x - E_{l,h})^2}{2\delta _{l,h} ^2}
                     \right \} \ ,
\label{eqn:spinDistribution}
\end{equation}
and the width, $\delta _{l,h}$, is propagated from the
kinetic energy distribution:
\begin{equation}
  \delta_{l,h} = \frac{\sigma _\mathrm{TKE}}
                     {\sqrt{E_l^2 + E_h^2}} E_{l,h} \ .
\end{equation}

The spin distribution, $R(J)$, is assumed to be proportional to the available spin states in the level density formula,
\begin{equation}
  R_{l,h}(J) = \frac{J+1/2}{f^2 \sigma^2_{l,h}(U)}
              \exp \left \{
                           -\frac{(J+1/2)^2}{2f^2 \sigma _{l,h}^2 (U)}
                   \right \} \ ,
\end{equation}
where $\sigma^2_{l,h}(U)$ is the spin cut-off parameter -- with adjustable scaling factor, $f$ -- and $U$ is the excitation energy corrected by the pairing energy.  The scaling factor $f$ is related to the scaling factor $\alpha$ used in our previous publications with \cgmf{} \cite{Becker2013,CGMF}.  After the population distribution is created for the heavy and light fragments, the statistical Hauser-Feshbach decay is performed for each fragment from the highest energy bin, typically $U_{l,h} + 2.5 \delta _{l,h}$, and the population of each residual nucleus is collected.

\subsection{Extension to multi-chance fission}
\label{sec:MCF}

Typically TKE behaves just like a monotonously decreasing function of the incident neutron energy~\cite{Madland2006}, except for a kink at low incident neutron energies that is seen in some experimental data sets, such as~\cite{Duke2015,ANLNDM64}. To account for this feature, TKE of the two fission fragments is parametrized by a piecewise linear function as
\begin{equation}
  \langle \mathrm{TKE} \rangle (E_\mathrm{inc}|m) =
    \begin{cases}
      a + b E_\mathrm{inc}, & \text{if } E_\mathrm{inc} \le E_0 \\
      c + d E_\mathrm{inc}, & \text{if } E_\mathrm{inc} \ge E_0
    \end{cases} \ ,
\label{eqn:TKEE}
\end{equation}
where $a$, $b$, $d$, and $E_0$ are fitting parameters, and $c$ is determined by the continuity at $E_0$,
\begin{equation}
  c = a + (b-d)E_0 \ .
\end{equation}

In principle, the TKE distribution in Eq.~(\ref{eqn:TKEE}) can be parametrized separately for each $A_c-m+1$ compound nucleus that can be created before the nucleus fissions. Since the average excitation energy causing fission, $\langle E_f \rangle (m)$, is also already calculated by \coh{}, an equivalent incident energy, $E_\mathrm{eq}$,
\begin{equation}
  E_\mathrm{eq}(m) = \langle E_f \rangle (m) - S_n (m) \ ,
  \label{eqn:Eeq}
\end{equation}
is substituted into Eq.~(\ref{eqn:TKEE}) to obtain TKE at each fission chance. This method allows us to compare the calculated TKE with the experimentally observed data by folding over the multi-chance fission probabilities with each TKE$(E_\mathrm{inc}|m)$ distribution,
\begin{eqnarray}
  \langle \mathrm{TKE} \rangle (m) \nonumber \\
  &=& \frac{\displaystyle\int \sigma_f(m,E_x) \mathrm{TKE}(E_x - S_n(m)) dE_x}
         {\displaystyle\int \sigma_f(m,E_x) dE_x} \nonumber \\
  &\simeq& \mathrm{TKE} (E_\mathrm{eq}|m) \ ,
\end{eqnarray}
and finally the average over the multi-chance contributions reads
\begin{equation}
  \langle \mathrm{TKE} \rangle (E_\mathrm{inc})
   = \sum_m \mathrm{TKE} (E_\mathrm{eq}|m) P_f(m) \ .
\end{equation}

To perform the multi-chance fission calculations, we first generate all of the fission pairs for the $A_c -m+1$ compound nucleus for each $m^\mathrm{th}$-chance fission channel energetically available, as from Eq.~(\ref{eqn:MCFprob}).  A more detailed discussion of how \beoh{} keeps track of the fission yields from multi-chance fission is found in Appendix \ref{app:MCFdetails}.  \beoh{} has the option to read in the yield parametrization from an external file.  The parameterization can be given for each chance fission independently, or, if these parametrizations are not given, \beoh{} will use the parametrization for first-chance fission, shifting any dependence on $A_c$ to $A_c -m+1$ and using the appropriate $E_\mathrm{eq}$ instead of $E_\mathrm{inc}$.  

\subsection{Calculation of prompt observables}

For each fission fragment, the statistical Hauser-Feshbach decay is calculated, providing results for the emitted neutrons and $\gamma$ rays as well as the residual fragment.  These results are then weighted with the initial fission yields to calculate the final observables.  For first-chance fission, this averaging is straightforward.  The calculated neutron spectra for the heavy and light fragments, $\phi_{l,h}(\epsilon)$, integrate to the average neutron multiplicity,
\begin{equation}
\int \phi _{l,h} (\epsilon) d\epsilon  = \overline{\nu}_{l,h},
\end{equation}
where $\overline{\nu}_k = \overline{\nu}_l + \overline{\nu}_h$ is the average total prompt number of neutrons for a particular fragment split, $k$.  The total energy for each fragment split is calculated from the sum of the light and heavy fragment spectra,
\begin{equation}
E_k = \int \epsilon \left \{ \phi _l (\epsilon) + \phi _h (\epsilon) \right \} d \epsilon,
\end{equation}
and the average energy of the emitted neutrons is 
\begin{equation}
\langle \epsilon \rangle _k = \frac{E_k}{\overline{\nu}_k}.
\end{equation}
The average total neutron multiplicity and average neutron energies are calculated by weighing each $\overline{\nu}_k$ and $\langle \epsilon \rangle _k$ with the corresponding fission yield.  


 
The calculations for these quantities when the multi-chance fission channels are open are slightly more involved.  Because the yields are not necessarily the same for the light and heavy fragments (see Appendix \ref{app:MCFdetails}), the prompt neutron multiplicity and average energy are now defined as 
\begin{equation}
\overline{\nu}_{\mathrm{sd}} = \sum _k \left [ Y_{l}(k) \overline{\nu}_l + Y_h (k) \overline{\nu}_h \right ] \ ,
\end{equation}
and
\begin{equation}
E _{\mathrm{sd}} = \sum _k \left [ Y_l(k) \overline{\nu}_l \langle \epsilon \rangle _l + Y_h(k) \overline{\nu}_h \langle \epsilon \rangle _h \right ] \ ,
\end{equation}
where the subscript ``sd'' stands for statistical decay --- to distinguish between the pre-fission, ``pf'', neutron observables, that is the average number of and average energy of the neutrons emitted before scission.  These are calculated as
\begin{equation}
 \overline{\nu}_{\mathrm{pf}} = \sum_m (m-1) P_f(m) \ ,
 \end{equation}
 and 
\begin{equation}
E_{\mathrm{pf}} = \sum_m (m-1) P_f(m) \langle \epsilon_{\mathrm{pf}} (m) \rangle \ .
\end{equation}
Then, the average total prompt neutron multiplicity and average neutron energies are
\begin{equation}
\overline{\nu} = \overline{\nu}_\mathrm{pf} + \overline{\nu} _\mathrm{sd},
\end{equation}
and 
\begin{equation}
  \langle \epsilon \rangle = E / \overline{\nu} \ ,
\end{equation}
where the total neutron energy, $E$, is
\begin{equation}
  E = E_{\mathrm{pf}} + E_{\mathrm{sd}} \ .
\end{equation} 
The total energy of the pre-fission neutrons, $E_{pf}$, is calculated by \coh{} and is included as an input into \beoh{}.  The $\gamma$-ray properties are calculated in much the same manner.

\subsection{Calculation of delayed observables}
\label{sec:delayedCalc}

Neutron emission, as described in the previous section, leads to the calculation of the independent fission product yields, $Y_I(A,Z)$.  At this step, we include an index, $M$, to explicitly keep track of the isomeric states, $Y_I(A,Z,M)$.  \beoh{} also calculates the cumulative fission product yields, $Y_C(A,Z,M)$ using decay data.  These calculations are performed in a time-independent manner, by setting $dY_C(A,Z,M)/dt=0$,
\begin{eqnarray}
  Y_C(A_i,Z_i,M_i)
  &=& Y_I(A_i,Z_i,M_i) \nonumber \\
  &+& \sum_j^N \sum_l^{L_j} Y_C(A_j,Z_j,M_j) b_{jl} \delta_{jl,i} \ ,
\end{eqnarray}

\noindent where $N$ is the total number of nuclei that produce the $i^\mathrm{th}$ nucleus, $(A_i,Z_i,M_i)$, and $\delta_{jl,i}$ connects $Y_C(A_j,Z_j,M_j)$ to $Y_C(A_i,Z_i,M_i)$ through any of the branching ratios, $b_{jl}$, with $L_j$ total decay modes.  The branching ratios are normalized as
\begin{equation}
\sum \limits _l ^{L_{j}} b_{jl} = 1 \ .
\end{equation}
The $\beta$ decay can be calculated directly \cite{Mumpower2016,Moller2019}, however, we use the ENDF/B-VIII.0 \cite{ENDFB8} decay data library for these branching ratios. In addition, the JENDL-4 decay library may be employed to study the impact of the decay data on the cumulative yields.

When the branching ratio of the $i^\mathrm{th}$ nucleus includes a neutron emission mode, the $\beta$-delayed neutron yield is calculated as 
\begin{equation}
\nu _{d,i} = Y_C(A_i,Z_i,M_i) b_{il} \left \{ A_l - A_i \right \} \ ,
\end{equation}
where $A_l - A_i$ is the mass difference between the parent and daughter fragments; this difference is typically one but is explicitly included to take into account the possibility of two-neutron emission.  Then the average total delayed neutron yield is calculated as
\begin{equation}
\overline {\nu}_d = \sum \limits _i \nu _{d,i} \ .
\end{equation}

\subsection{Constraining the yield parametrization}
\label{sec:fitting}


The initial parametrizations for $Y(A)$ and $\langle \mathrm{TKE} \rangle (E_\mathrm{inc})$ are based on the parametrization used by \cgmf{} \cite{CGMF,Becker2013} but have been re-parametrized and refit in many cases.  The mass yield parametrization is constrained separately for $^{235}$U(n,f), $^{234}$U(n,f), and $^{233}$U(n,f), as a function of incident neutron energy.  Around incident energies of 19--20 MeV, when the fourth-chance fission channel opens, the $Y(A|E_\mathrm{inc})$ parametrization of $^{233}$U(n,f) is used, with $A_c=234$ shifted to $A_c-1=233$, as experimental data for $^{232}$U(n,f) are not readily available.  

The \beoh{} parametrization for $\mathrm{TKE}(A_h)$ and $\sigma _\mathrm{TKE} (A_h)$ in Eqs. (\ref{eqn:TKEA}) and (\ref{eqn:sigmaTKE}) has been changed from \cgmf{} (a sum of polynomials) but was fit to the \cgmf{} parametrization for $^{235}$U(n,f) at thermal, which has been constrained by experimental data.  The same parametrization for the $^{235}$U, $^{234}$U, and $^{233}$U compounds is used as for the $^{236}$U compound.  In addition, the magnitude of the fitted $\sigma_\mathrm{TKE}(A_h)$ is typically reduced in comparison to experiment to better reproduce experimentally measured neutron multiplicity distributions.

Finally, the spin cut-off parameter and the charge distribution factors were tuned to the chain yield data from the England and Rider evaluation \cite{EnglandRider} with the lowest listed uncertainties, $A=$ 132, 133, 135, 136, 143, 144, 145, 145, 148, along with the prompt and delayed $\overline{\nu}$ from ENDF/B-VIII.0~\cite{ENDFB8}, using a Kalman filter \cite{Kalman1960}.  The parameters used in this work are listed in Table \ref{tab:parameters} in Appendix \ref{app:parmValues}.

Ultimately, the goal of this work is to show the viability of the updates to \beoh{} to take into account multi-chance fission, not necessarily to best optimize each of the input yield models.  More thorough optimizations are being performed and will be discussed in more detail in future work, along with the extension to major actinides beyond $^{235}$U, which is the only target shown here.

\section{Results and Discussion}
\label{sec:results}



\begin{figure}
\centering
\includegraphics[width=0.5\textwidth]{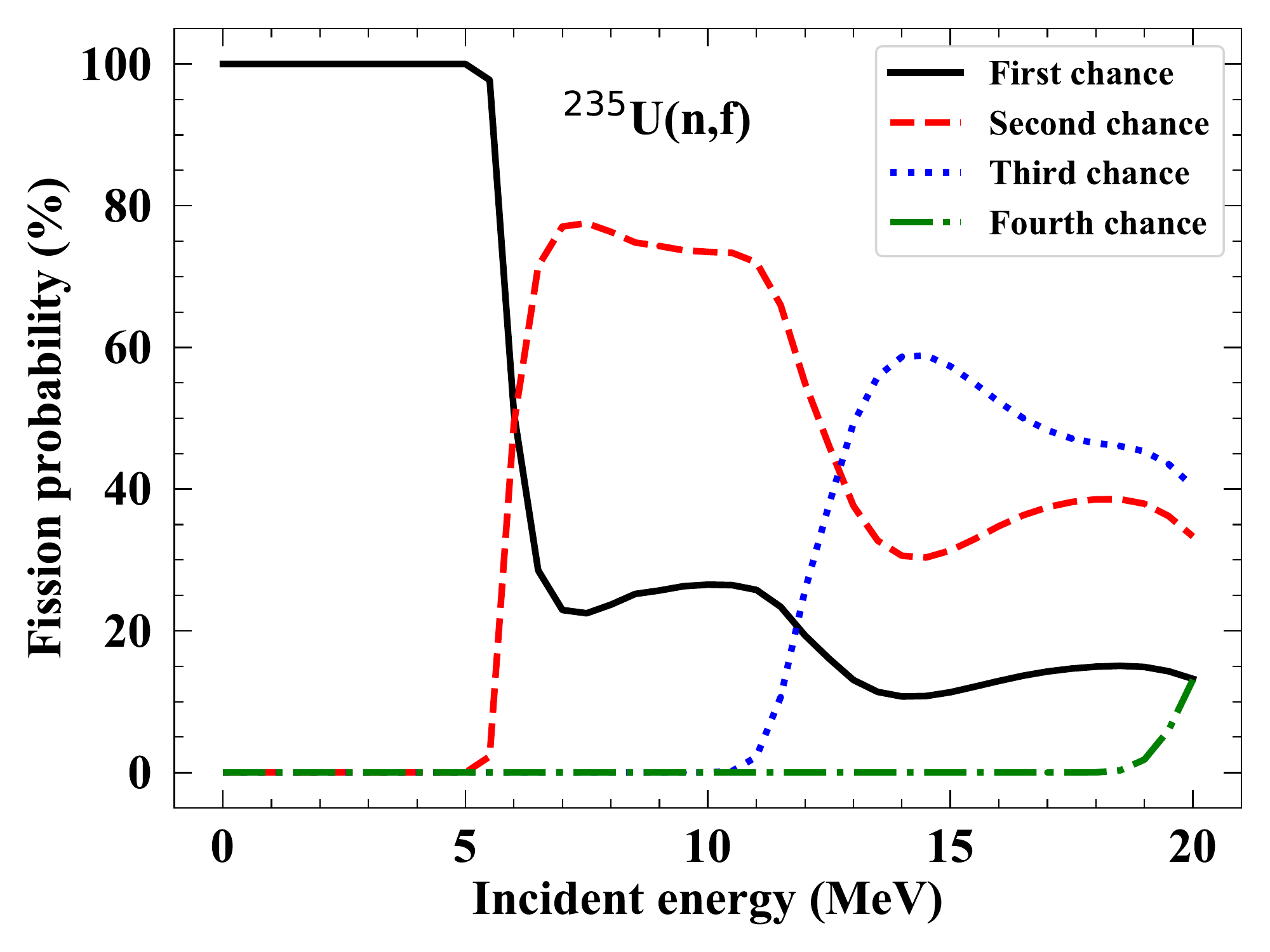}
\caption{Multi-chance fission probabilities, in percentages, for $^{235}$U(n,f) for first- (black solid), second- (red dashed), third- (blue dotted), and fourth- (green dash dotted) chance fission, as calculated by \coh{}.}
\label{fig:MCFprob}
\end{figure}

\begin{figure}
\centering
\includegraphics[width=0.5\textwidth]{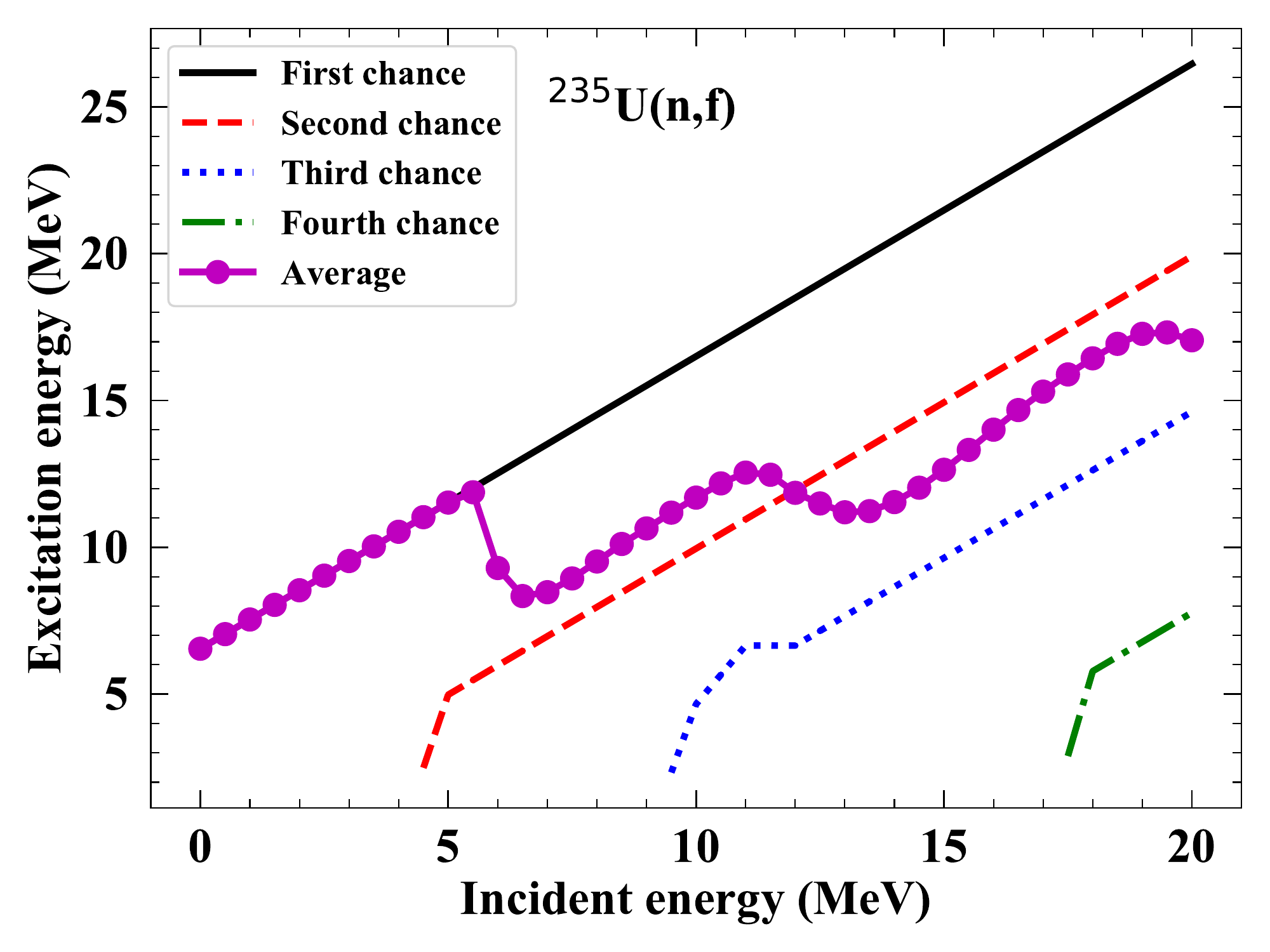}
\caption{Average excitation energy causing fission for first- (black solid), second- (red dashed), third- (blue dotted), and fourth- (green dash dotted) chance fission for $^{235}$U(n,f) as calculated by \coh{}.  The magenta dots show the average -- the sum of each $m^\mathrm{th}$-chance excitation energy folded with the $m^\mathrm{th}$-chance fission probability, as shown in Fig. \ref{fig:MCFprob}.}
\label{fig:Eexc}
\end{figure}

We consider, as an example for this multi-chance fission extension, $^{235}$U(n,f) with incident energies from thermal to 20 MeV.  As discussed in Sec. \ref{sec:theory}, there are several quantities that must be pre-calculated and used as input to \beoh{}.  The multi-chance fission probabilities calculated from \coh{} are shown in Fig. \ref{fig:MCFprob}, and the average excitation energy causing fission is shown in Fig. \ref{fig:Eexc}.  In Fig. \ref{fig:Eexc}, we note that even though the excitation energy for each multi-chance fission channel increases essentially linearly, the average excitation energy causing fission does not increase as sharply, as the neutron emitted before fission removes energy, cooling the residual compound system.  

\begin{figure}
\centering
\includegraphics[width=0.5\textwidth]{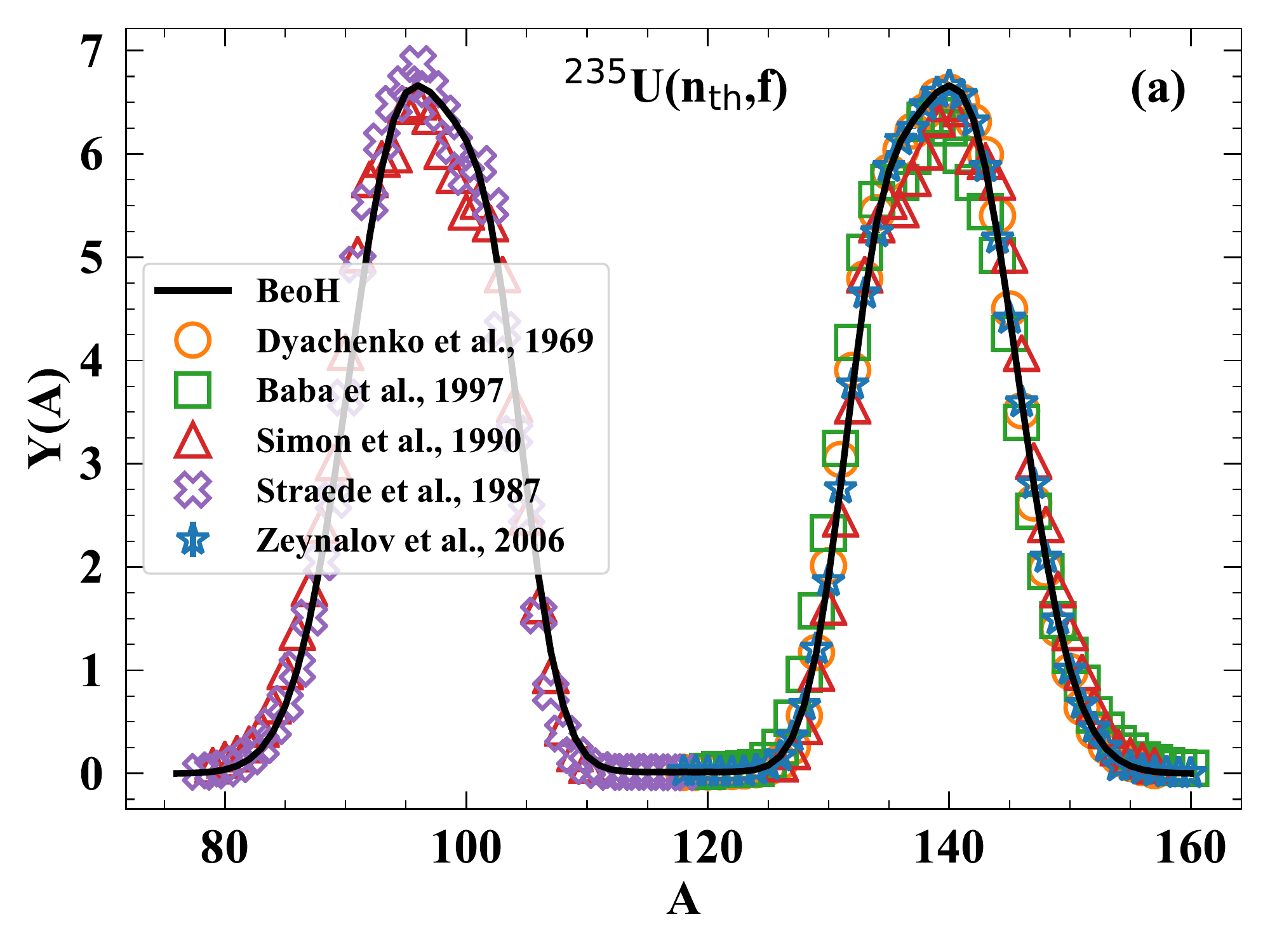} \\
\includegraphics[width=0.5\textwidth]{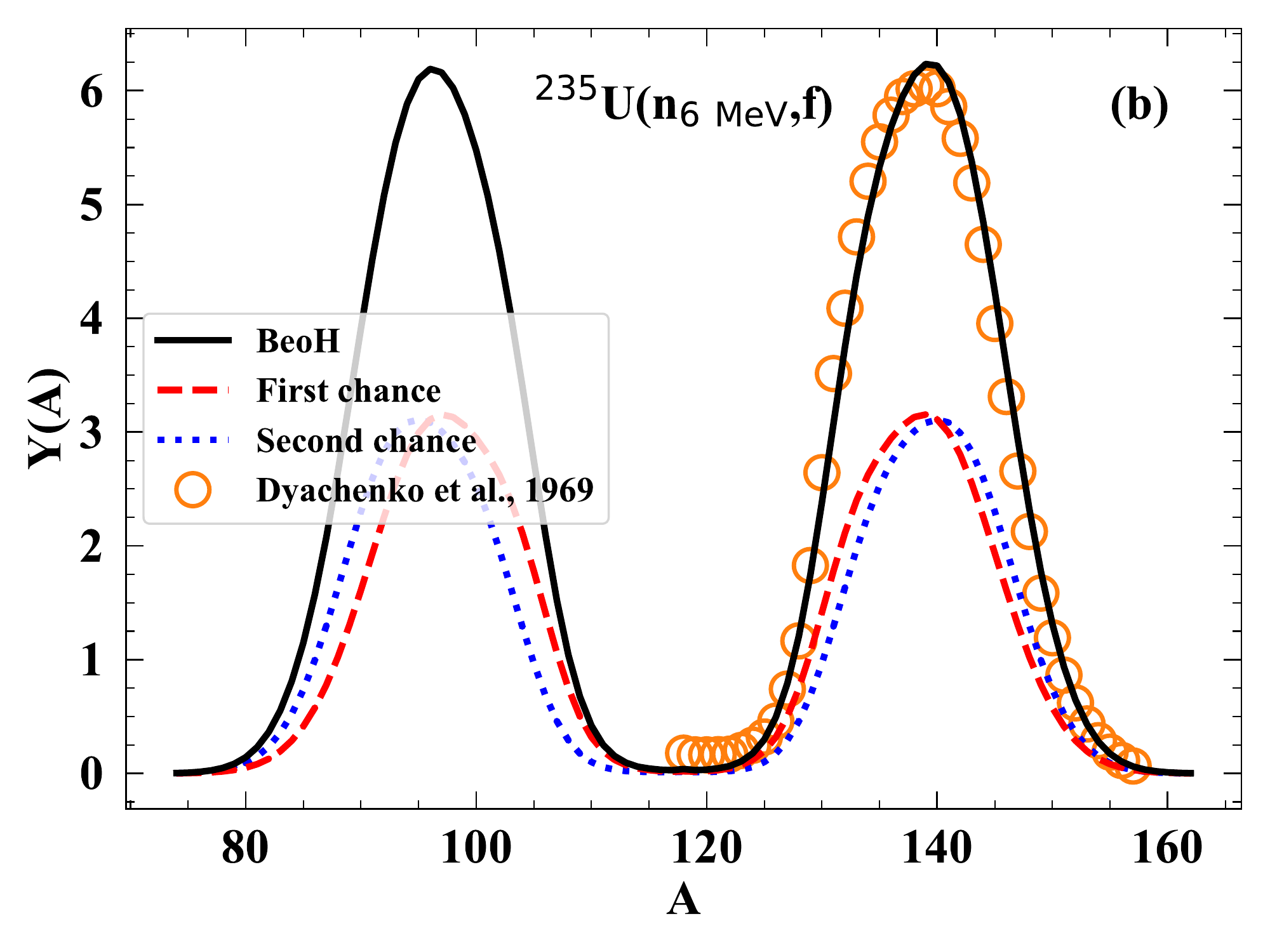} \\
\includegraphics[width=0.5\textwidth]{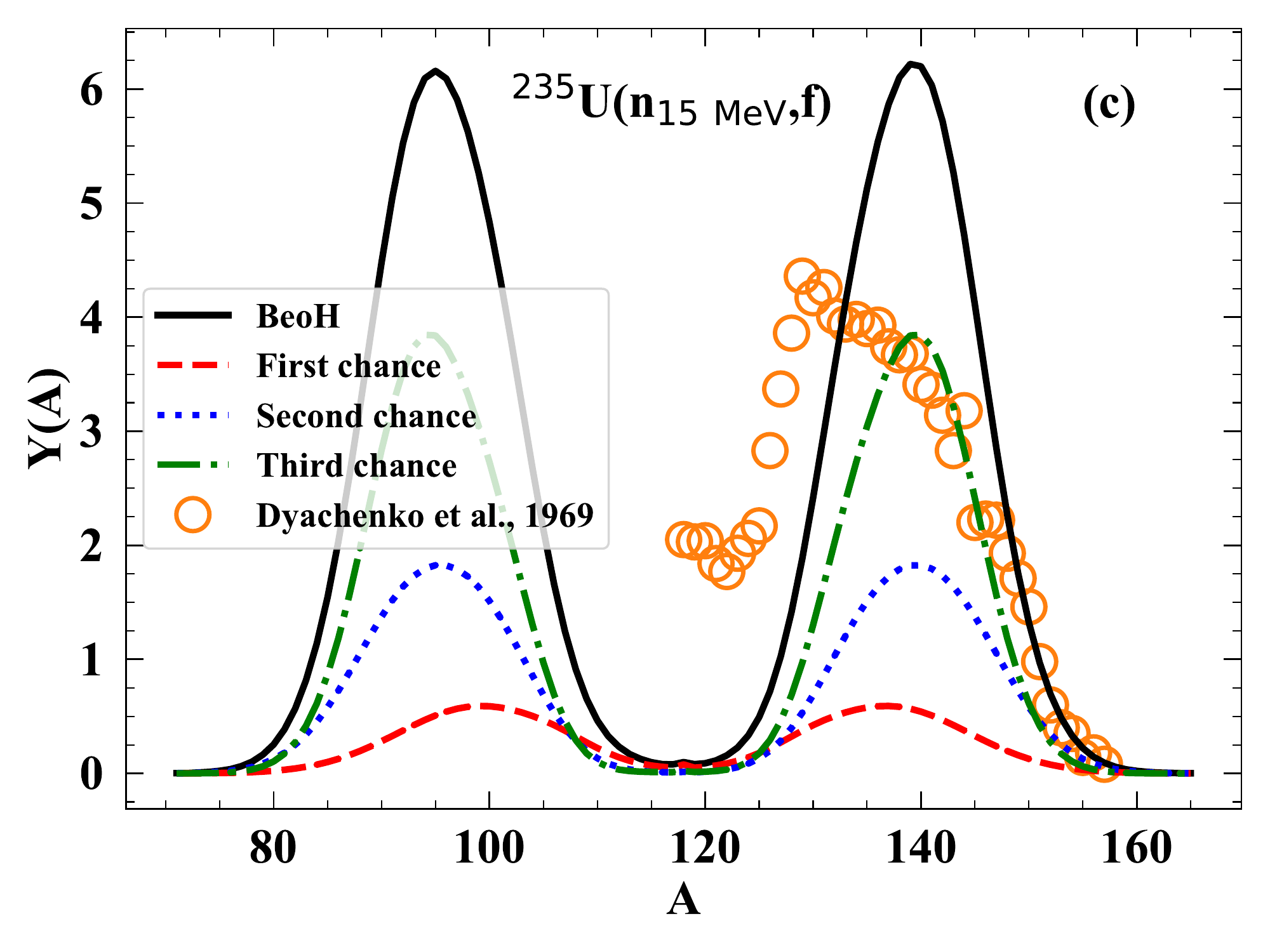} \\
\caption{Black solid, pre-neutron emission mass yields for $^{235}$U(n,f) with incident neutron energies of (a) thermal, (b) 6 MeV, and (c) 15 MeV, compared to experimental data \cite{Dyachenko1969,Baba1997,Simon1990,Straede1987,Zeynalov2006}.  Red dashed lines represent the part of the distribution coming from first-chance fission, blue dotted lines for second-chance fission, and green dot-dashed lines for third-chance fission.}
\label{fig:YA}
\end{figure}

In addition to the fission probabilities and excitation energies from \coh{}, we also include the mass yield parametrization, kinetic energy parametrization, and excitation energy sharing as a function of mass in external files.  The optimization for these parametrizations is described Sec. \ref{sec:fitting}, and here, we show how the multi-chance parameterization compares to experimental data as a function of incident neutron energy.  First, in Fig. \ref{fig:YA}, we show the pre-neutron emission mass yields at (a) thermal, (b) 6-MeV, and (c) 15-MeV incident neutron energies compared to experimental data.  As the incident neutron energy increases, the shape of the mass yield distribution changes smoothly, and even at $E_\mathrm{inc}=6$ MeV in panel (b), where first-chance fission and second-chance fission are taken to be essentially equal (and therefore the parametrization for $Y(A)$ is taken to be a combination of fission of the $^{236}$U and $^{235}$U compounds), the \beoh{} calculation reproduces the shape of the data reasonably well.  As the incident energy increases further and the third-chance fission channel opens, as in panel (c), the experimental $Y(A)$ distribution is shifted and wider compared to the distribution from \beoh{}.  Although this discrepancy could indicate that the fission probabilities or the $Y(A|E_\mathrm{inc})$ parametrizations of the $^{235}$U and $^{234}$U fissioning compounds need to be adjusted, all experimental methods to determine the pre-neutron emission mass distribution measure the fragment masses after neutron emission and reconstruct the pre-neutron masses based on a determination of $\overline{\nu}(A)$; the lack of experimental data on $\overline{\nu}(A)$ at $E_\mathrm{inc}=15$ MeV could lead to larger uncertainties on the data than shown in Fig. \ref{fig:YA}(c).

\begin{figure}
\centering
\includegraphics[width=0.5\textwidth]{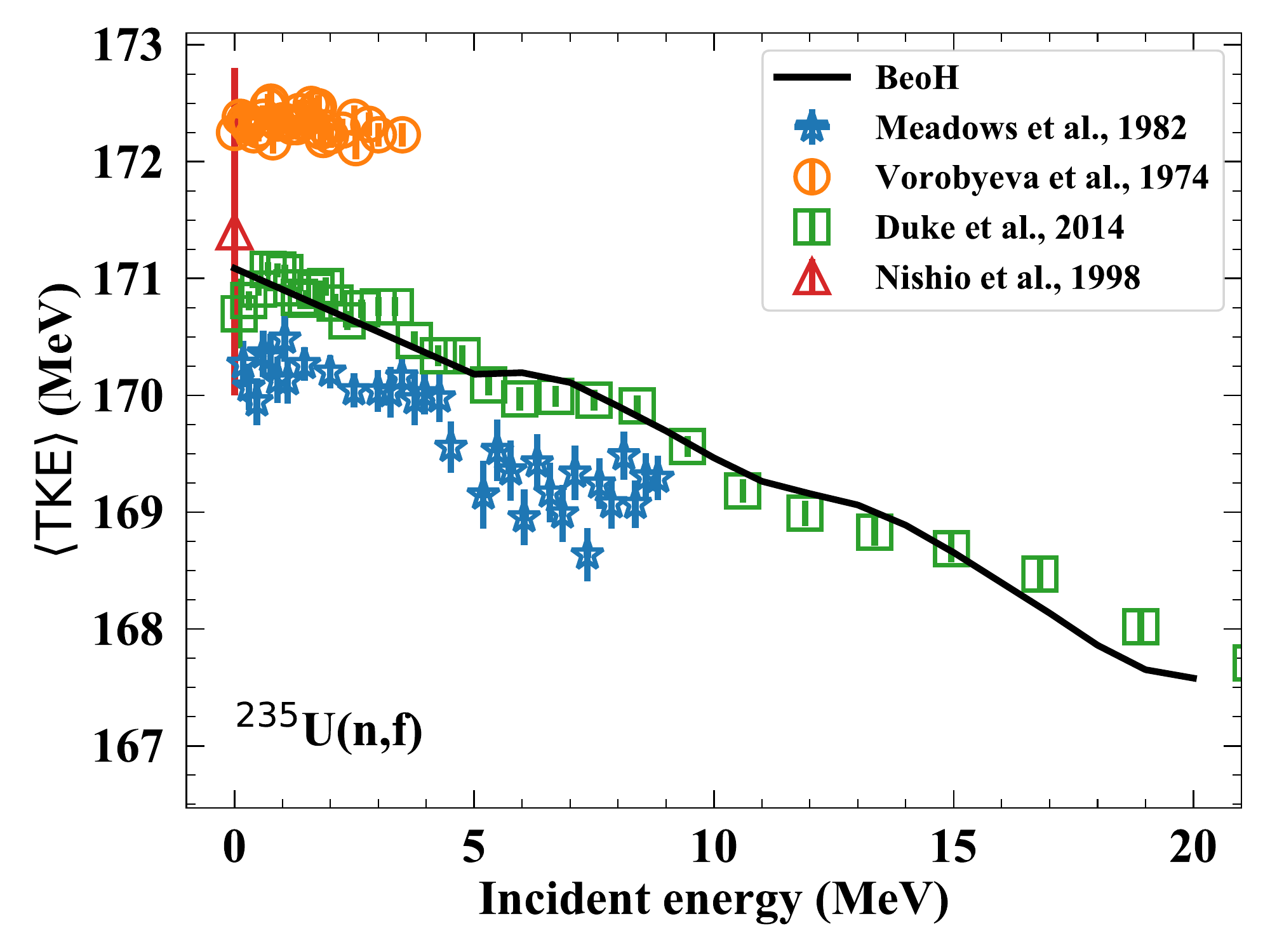}
\caption{Total kinetic energy as a function of incident energy for \beoh{} (solid) compared to experimental data \cite{ANLNDM64,Vorobyeva1974,Duke2015,Nishio1998} for $^{235}$U(n,f).}
\label{fig:TKE}
\end{figure}

In Fig. \ref{fig:TKE}, we show the average total kinetic energy as a function of the incident neutron energy, compared again to experimental data.  We can see that at the opening of the multi-chance fission channels there are kinks in the TKE, which come both from the slope change in the parametrization of $\langle \mathrm{TKE} \rangle$ in Eq. (\ref{eqn:TKEE}) and the averaging of TKE for the $m^\mathrm{th}$-chance fission compound with the multi-chance fission probabilities.  As there can be a lack of data for the lighter compound nuclei created by multi-chance fission, such as measurements for $\langle \mathrm{TKE} \rangle (E_\mathrm{inc})$ for $^{232}$U(n,f), comparison with $\langle \mathrm{TKE} \rangle$ data for $^{235}$U(n,f) -- as well as for $\overline{\nu}$ which is strongly anti-correlated to $\langle \mathrm{TKE} \rangle$ -- can help constrain these parameterizations for the other compounds.  Although our parametrization for $\langle \mathrm{TKE} \rangle (E_\mathrm{inc})$ in Eq.~(\ref{eqn:TKEE}) allows for a slope change at some incident energy $E_0$, here we have taken $E_0=0$ MeV for each compound.  The impact of this choice on the prompt and delayed observables is under further investigation.

\begin{figure}
\centering
\includegraphics[width=0.5\textwidth]{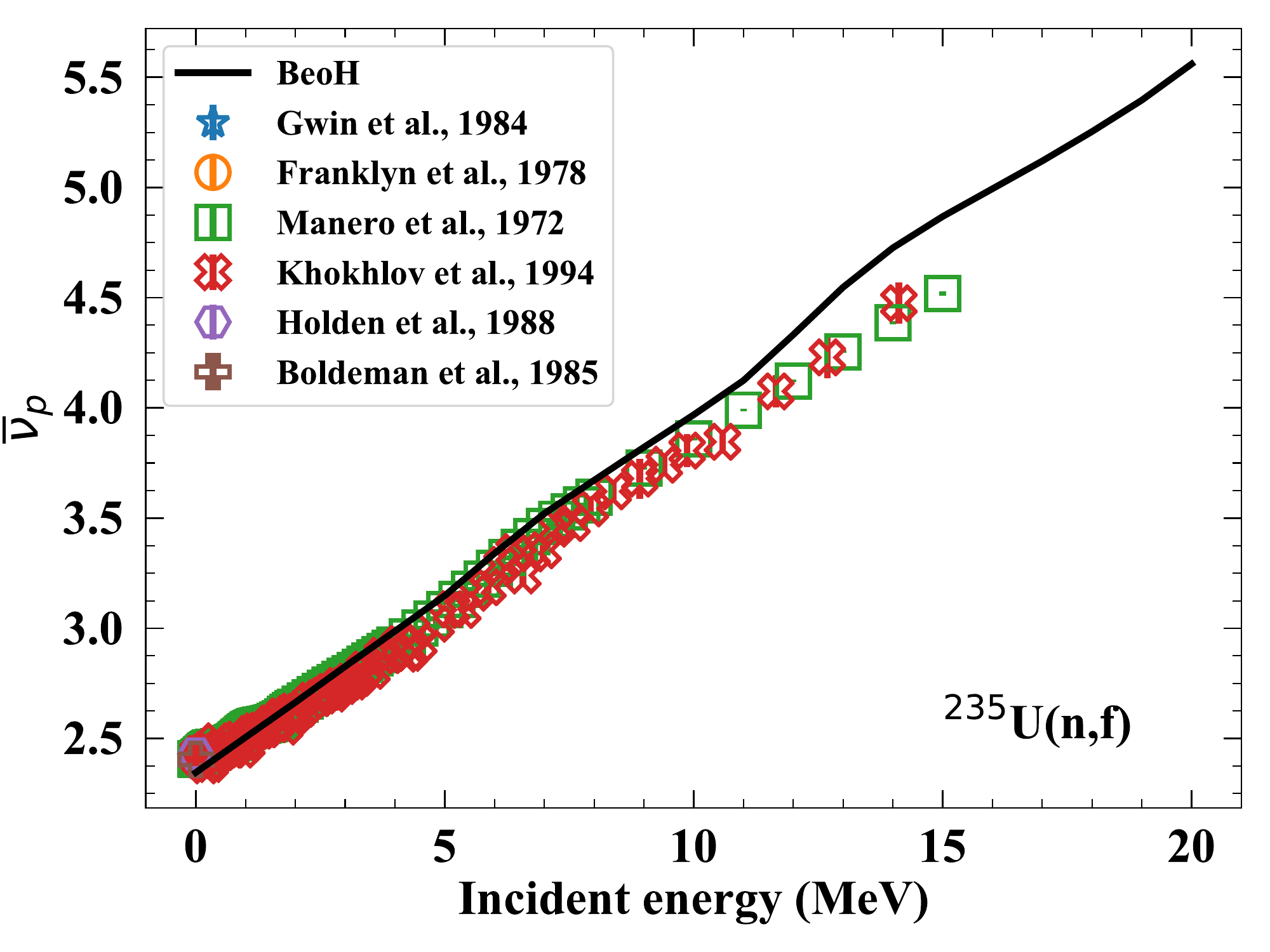}
\caption{Average prompt neutron multiplicity as a function of incident energy for \beoh{} (solid) compared to experimental data \cite{Gwin1986,Franklyn1978,Manero1972,Khokhlov1994,Frehaut1982,Holden1988,Boldeman1985} for $^{235}$U(n,f).}
\label{fig:nubar}
\end{figure}

We then show $\overline{\nu}$ as a function of the incident neutron energy in Fig. \ref{fig:nubar}.  The spread in the data is well reproduced by the \beoh{} calculation up to the opening of the second-chance fission channel, although the calculated $\overline{\nu}$ rises more quickly than the data.  The slope of $\overline{\nu}$ is a direct result of the slope of $\langle \mathrm{TKE} \rangle$ but could also be improved with the introduction of an energy-dependent spin cut-off factor.  Typically, $\overline{\nu}_p$ is assumed to be a straight line, particularly in nuclear data evaluations such as \cite{ENDFB8}.  However, our calculations show slight kinks in $\overline{\nu}_p$ at the openings of the multi-chance fission channels, similar to those seen in $\langle \mathrm{TKE} \rangle $ in Fig. \ref{fig:TKE}.

\begin{figure}
\centering
\includegraphics[width=0.5\textwidth]{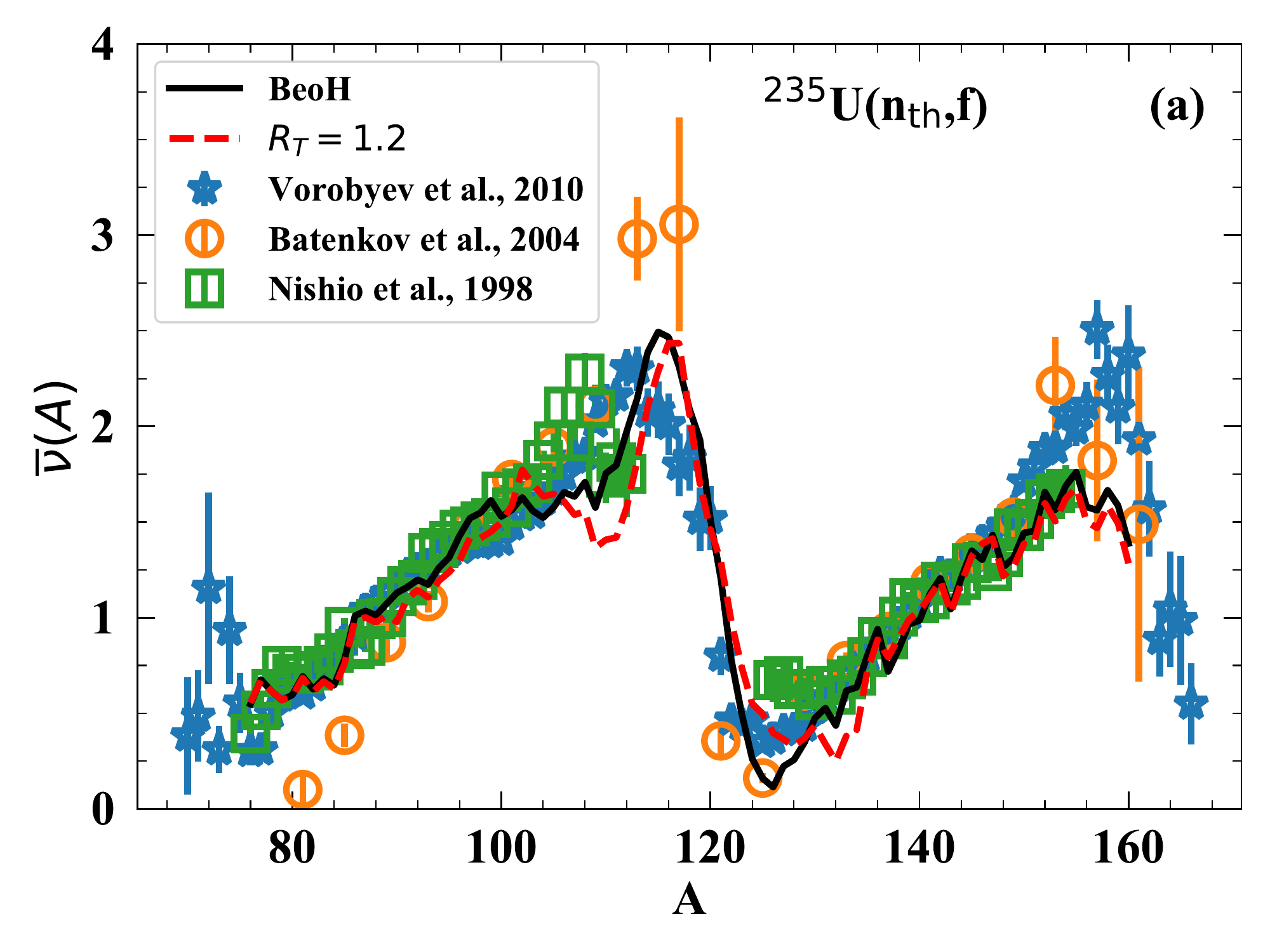} \\
\includegraphics[width=0.5\textwidth]{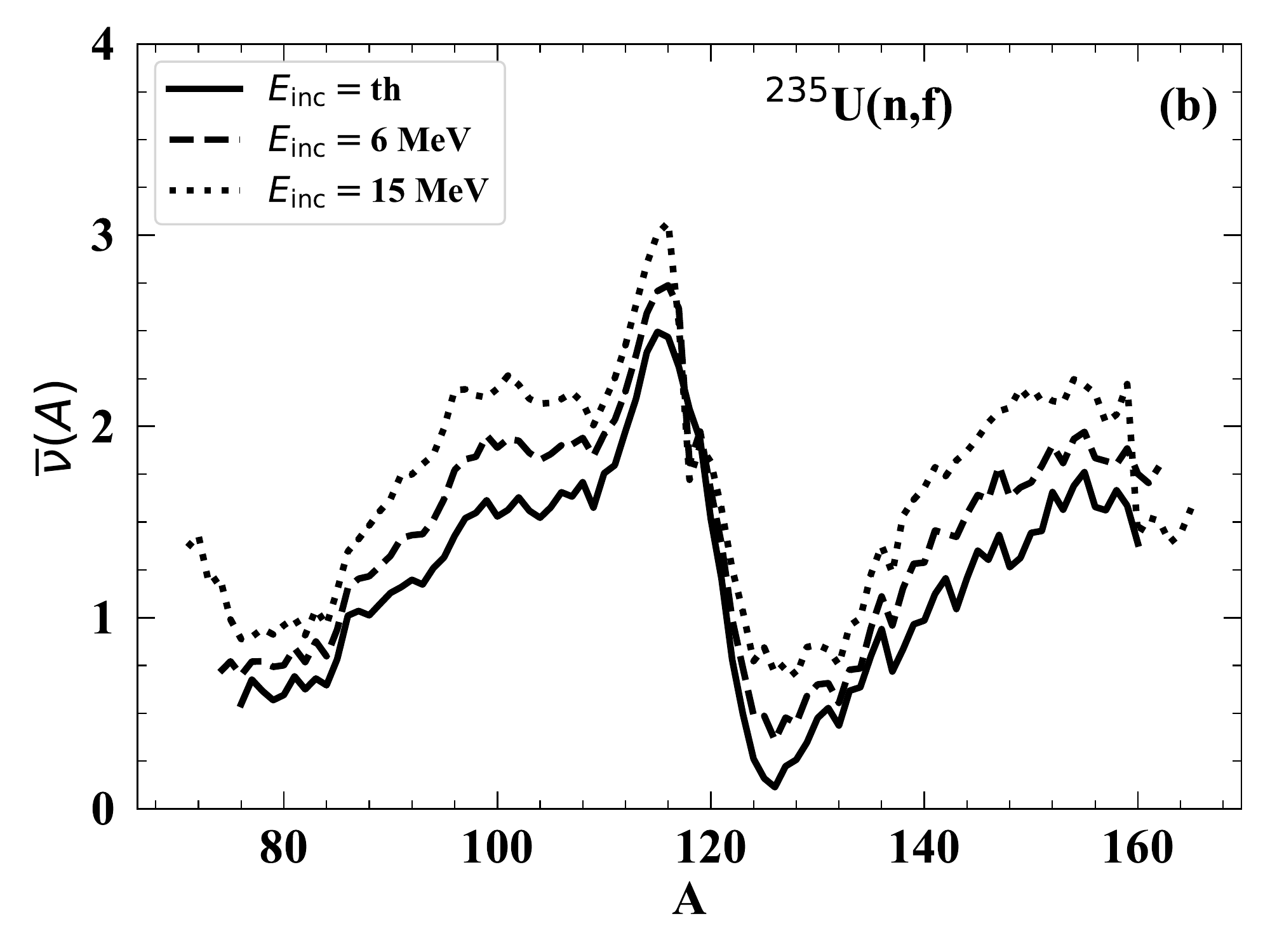} 
\caption{(a) Average neutron multiplicity as a function of mass for \beoh{} with $R_T$ as a function of mass (solid) and with fixed $R_T=1.2$ (dashed) compared to experimental data \cite{Vorobyev2010,Batenkov2005,Nishio1998} for $^{235}$U(n,f) for thermal incident neutrons.  (b) Average neutron multiplicity as a function of mass for thermal incident neutrons (solid), incident neutrons at 6 MeV (dashed), and incident neutrons at 15 MeV (dotted).}
\label{fig:nubarA}
\end{figure}

The excitation energy is shared between the heavy and light fragment through the $R_T$ parameter which has been tuned as a function of mass number.  The resulting average neutron multiplicity as a function of the fission fragment mass is shown in Fig. \ref{fig:nubarA}(a) by the solid curve.  The $A$-dependent parametrization of $R_T$ (solid curve) is compared to a fixed value of $R_T=1.2$ (dashed curve).  There are only slight changes between the two calculations of $\overline{\nu}(A)$ for $^{235}$U(n,f), except around the symmetric mass region.  Even though the mass-dependent $R_T(A)$ values have, on average, a less than 10\% difference from $R_T=1.2$, the largest differences are in the symmetric mass region and can be up to 30\%, leading to noticeable differences in $\overline{\nu}(A)$.  In both cases, there is a hint of a downturn in $\overline{\nu}(A)$, similar to the data of Vorobyev \emph{et al.} \cite{Vorobyev2010}.  

When multi-chance fission channels open, the values of $R_T(A)$ do not change from those at thermal incident energies unless explicitly specified (none currently for $^{235}$U(n,f)), which results in a rise of $\overline{\nu}(A)$ for all masses as the incident energy increase, Fig. \ref{fig:nubarA}(b).   Some experimental data \cite{Naqvi1986,Muller1984} and microscopic calculations \cite{Bulgac2019,Bulgac2020} indicate that as the incident neutron energy increases, more energy is given to the heavy fragment, increasing $\overline{\nu}$ for those fragments while $\overline{\nu}$ of the light fragments remains constant, consistent with phase-space arguments.  However, the available experimental data is still limited, so we leave the implementation of an energy-dependent $R_T(A)$ for future studies.  

\begin{figure}
\centering
\includegraphics[width=0.5\textwidth]{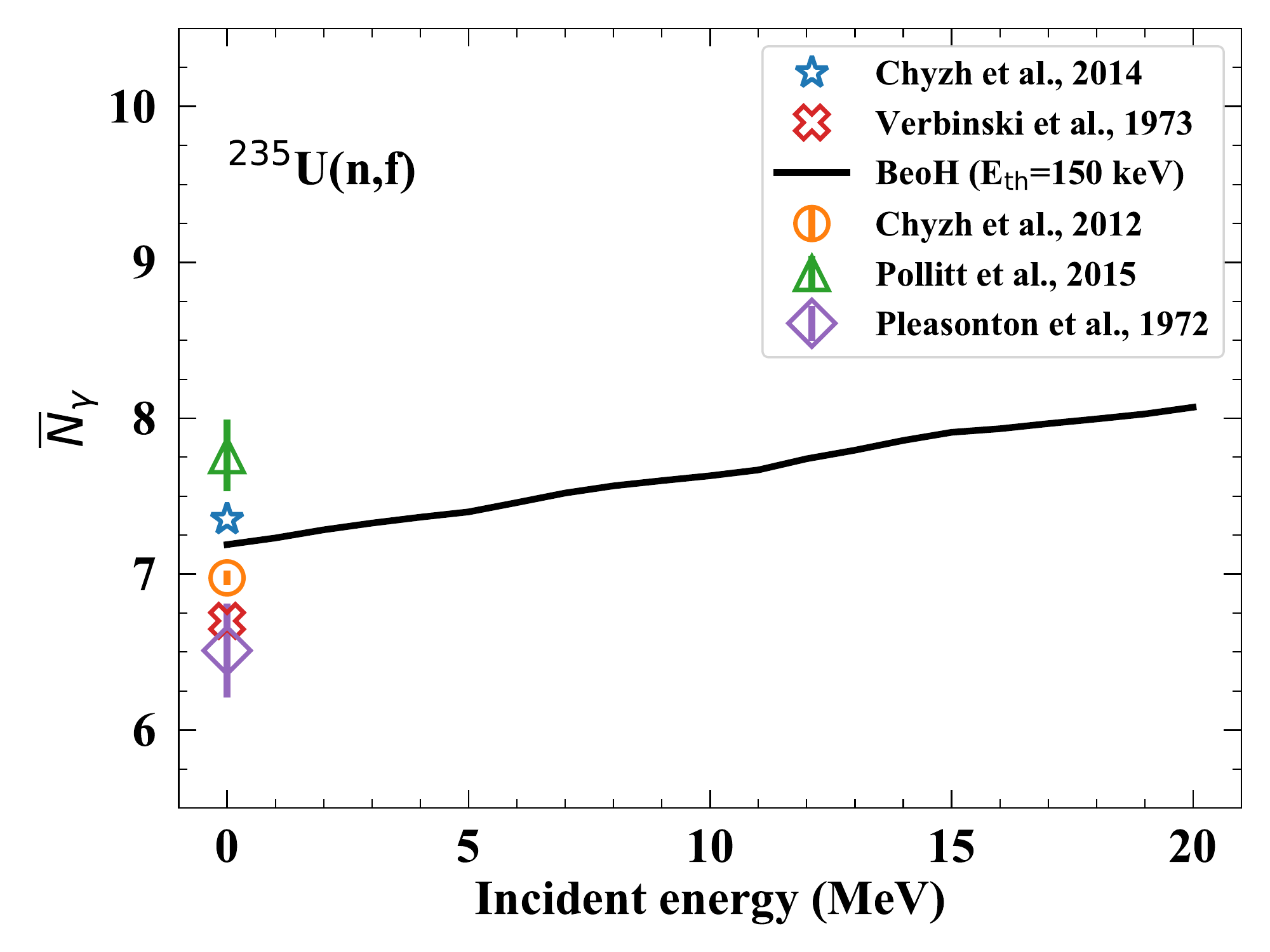}
\caption{Average $\gamma$-ray multiplicity as a function of incident energy for \beoh{} (solid) compared to experimental data \cite{Chyzh2014,Chyzh2012,Pollitt2015,Verbinski1973,Pleasonton1972}.  A lower bound cut-off on the outgoing $\gamma$-ray energy of $E_\mathrm{th}=150$ keV is included on the \beoh{} results, consistent with many experimental measurements.}
\label{fig:nugbar}
\end{figure}

In the same way, the properties of the prompt $\gamma$ rays are calculated.  As one example, in Fig. \ref{fig:nugbar}, we show the average $\gamma$-ray multiplicity, $\overline{N}_\gamma$, as a function of incident energy calculated from \beoh{} compared to experimental data at the thermal energy.  We include a threshold energy on the outgoing $\gamma$ rays, $E_\mathrm{th}$, of 150 keV, which is consistent with many experimental lower bounds.  Without the energy threshold included, the thermal value from \beoh{} is in good agreement with the thermal value from the ENDF/B-VIII.0 evaluation, which has been corrected for the experimental threshold energies of each data set.  The rise in $\overline{N}_\gamma$ as the incident energy increases can be changed by including an energy dependence in the spin cut-off factor, $f$.  This energy dependence was not included here and requires further studies due to the strong correlations between $f$, $\langle \mathrm{TKE} \rangle$, $\overline{\nu}_p$, and $\overline{N}_\gamma$.


\begin{figure}
\centering
\includegraphics[width=0.5\textwidth]{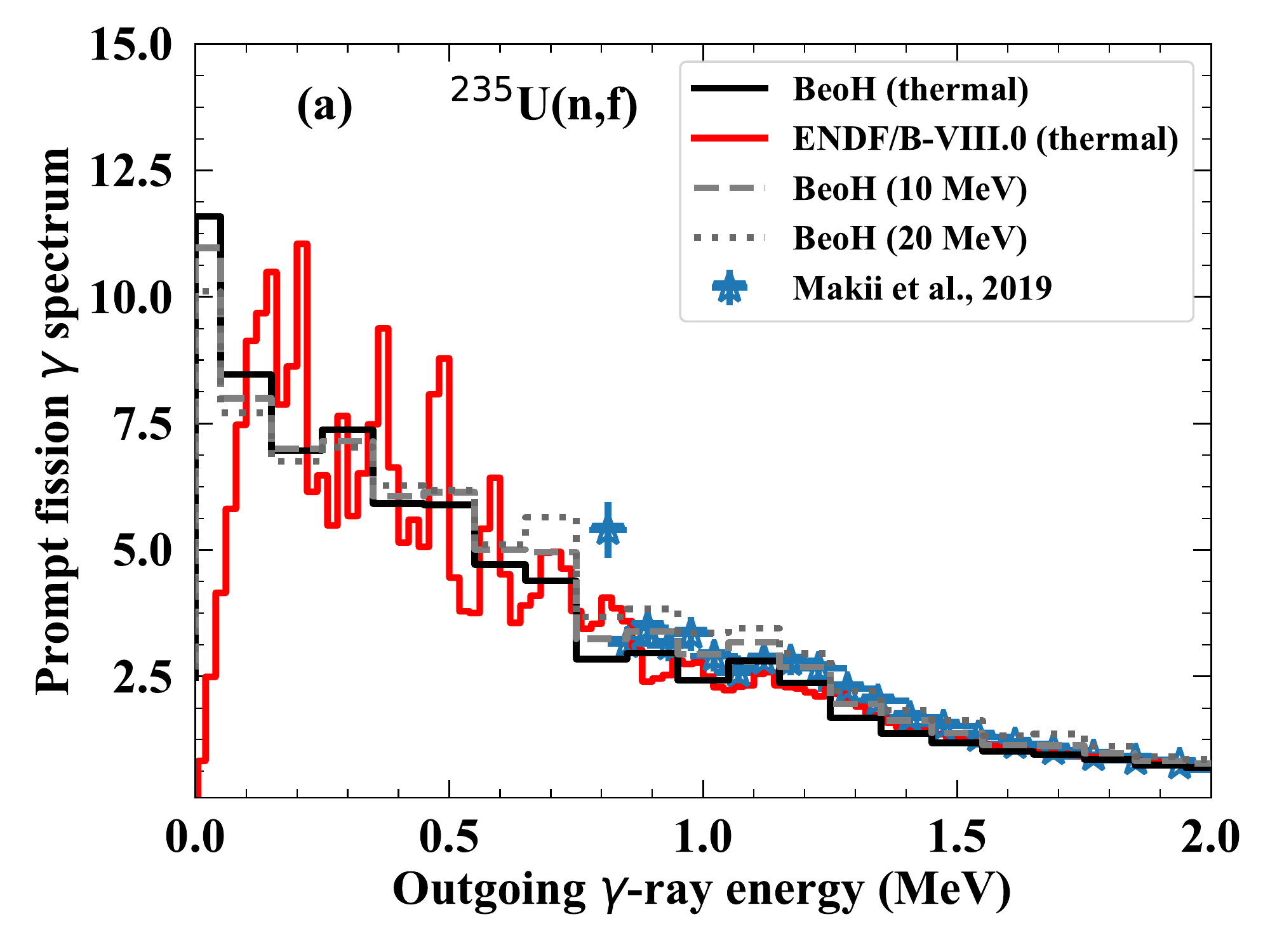} \\
\includegraphics[width=0.5\textwidth]{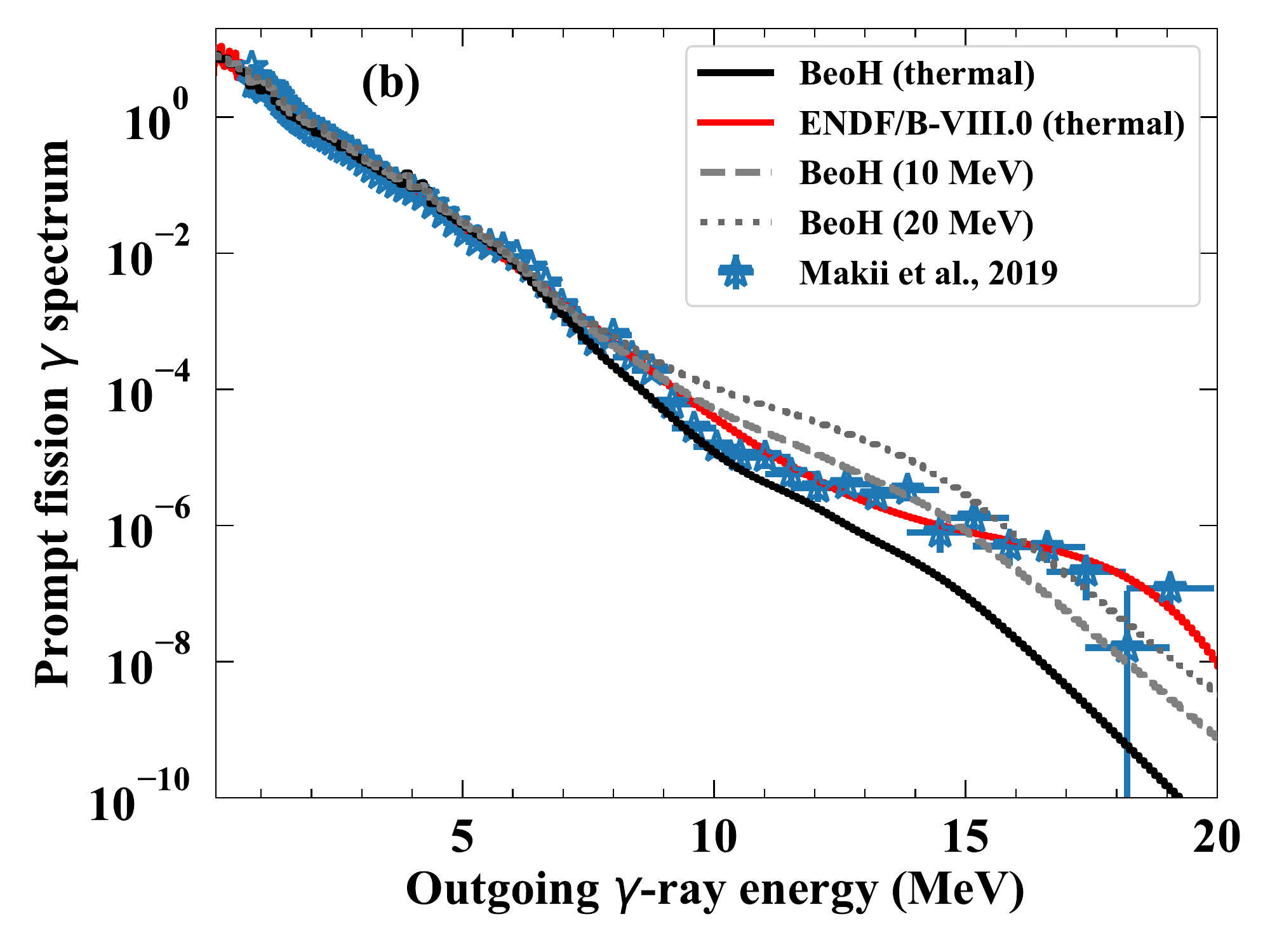} 
\caption{Prompt fission $\gamma$-ray spectrum for $^{235}$U(n,f) for outgoing $\gamma$-ray energies (a) up to 2 MeV and (b) up to 20 MeV.  Comparison between \beoh{} at thermal (black solid), $E_\mathrm{inc}=10$ MeV (grey dashed) and $E_\mathrm{inc}=20$ MeV (grey dotted), the ENDF/B-VIII.0 evaluation (red), and recent data by Makii, \emph{et al.}~\cite{Makii2019}.}
\label{fig:PFGS}
\end{figure}

We also plot the prompt fission $\gamma$ spectrum, PFGS, in Fig. \ref{fig:PFGS} focusing both on the discrete range of the spectrum in panel (a) and the high energy tail in panel (b).  In Fig. \ref{fig:PFGS}(a), the agreement between the shape of \beoh{} and the data and ENDF/B-VIII.0 evaluation is very good, although thousands of direct $\gamma$ lines are produced by \beoh{}, not all of the discrete levels in ENDF/B-VIII.0 are shown in the calculation, due to the binning in the outgoing $\gamma$-ray energy for better visibility.  As the incident energy of the \beoh{} calculation is increased, there is not much difference seen in the calculations, besides the magnitudes of the levels changing.  In the high-energy tail in Fig. \ref{fig:PFGS}(b), we see that the shape of the PFGS no longer follows the the ENDF/B-VIII.0 evaluation or experimental data above 10 MeV outgoing $\gamma$-ray energy \cite{Makii2019}.  However, contrary to the evaluation, which does not contain an energy-dependence in the PFGS, we see an increase in the tail of the PFGS from \beoh{} as the incident neutron energy increases.  This feature also leads to a slight increase in the average $\gamma$-ray energy between thermal and 20 MeV incident neutron energies (about 100 keV).

\begin{figure*}
\centering
\begin{tabular}{ccc}
\includegraphics[width=0.33\textwidth]{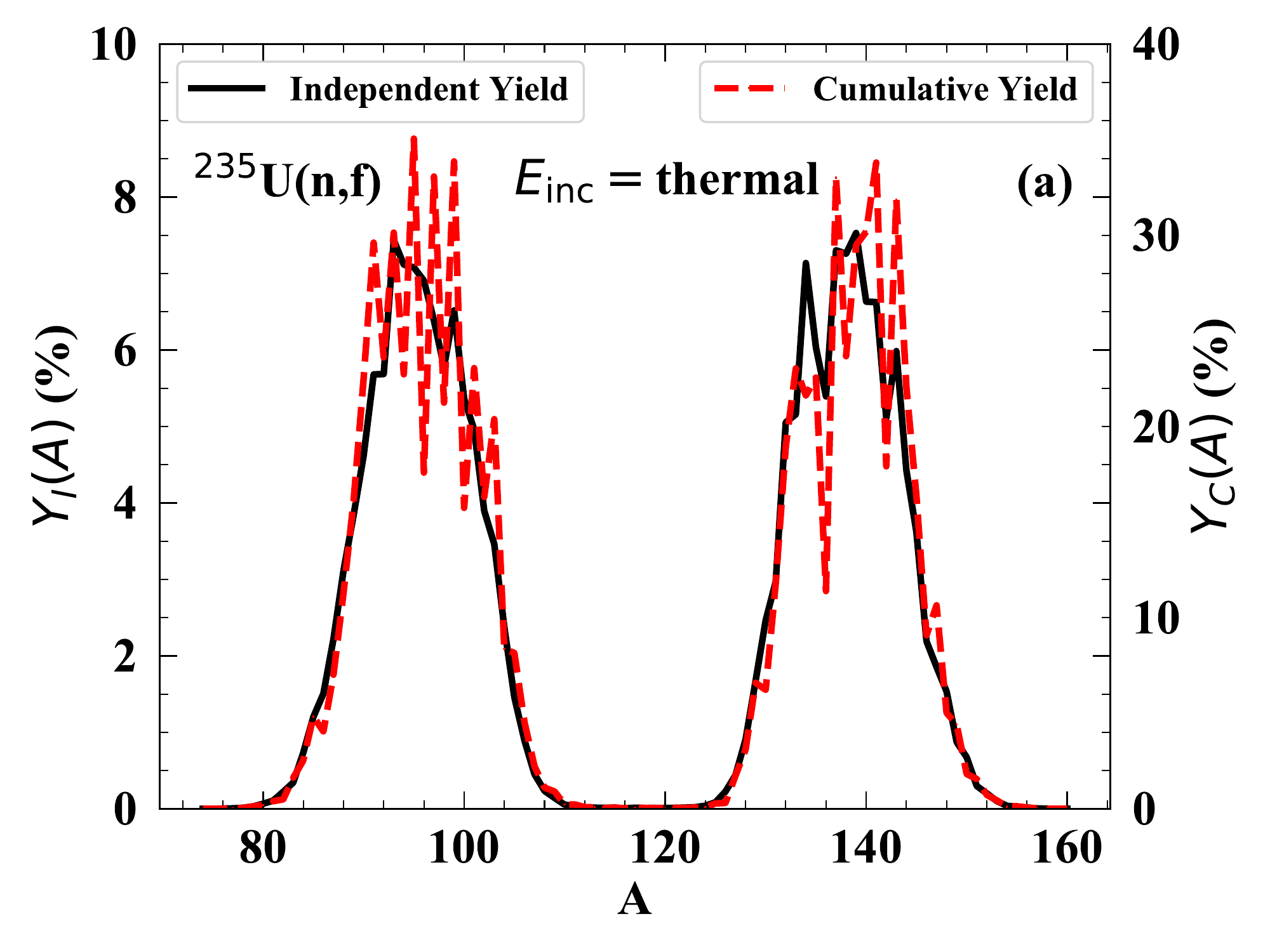} & \includegraphics[width=0.33\textwidth]{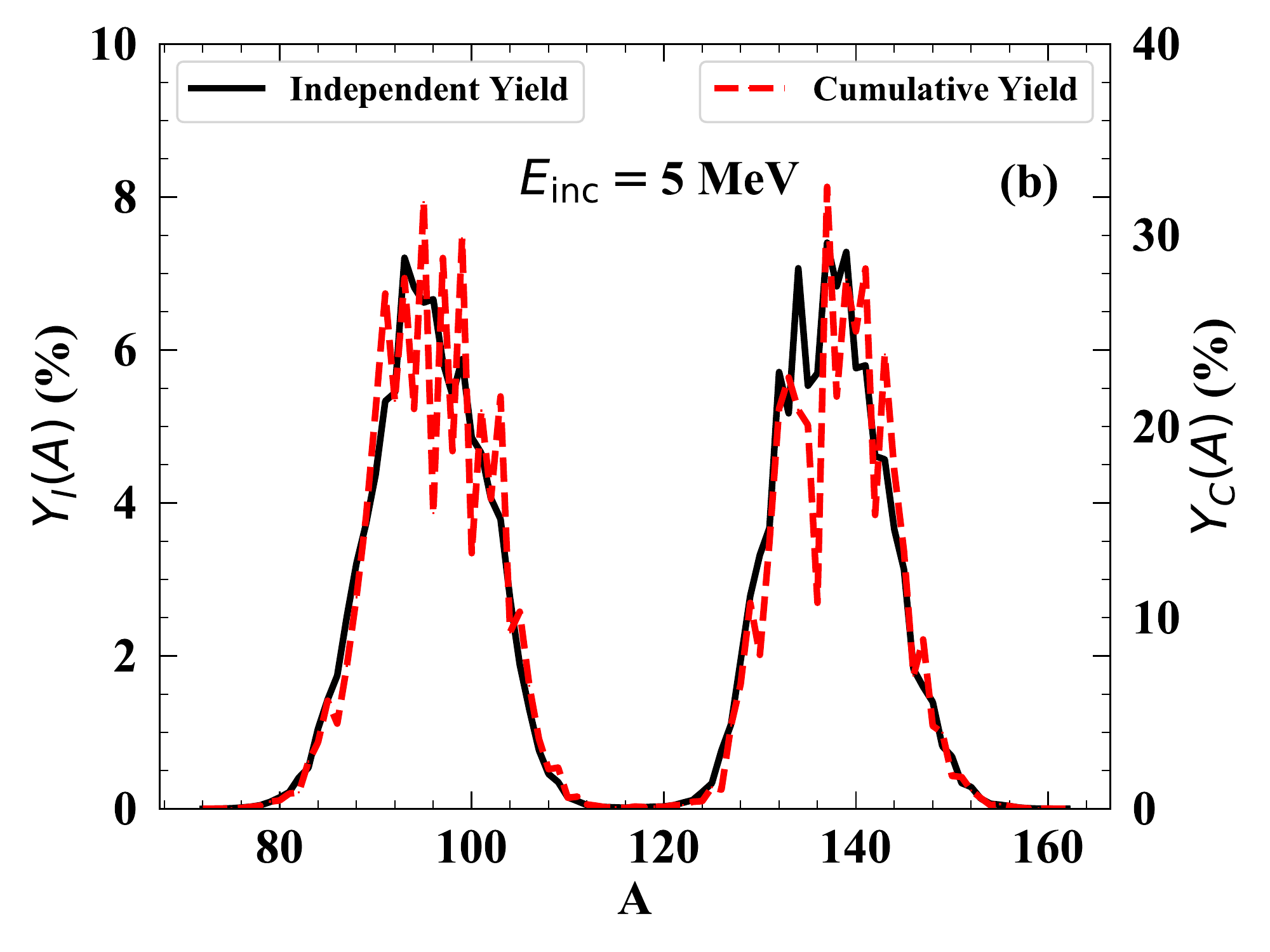} & \includegraphics[width=0.33\textwidth]{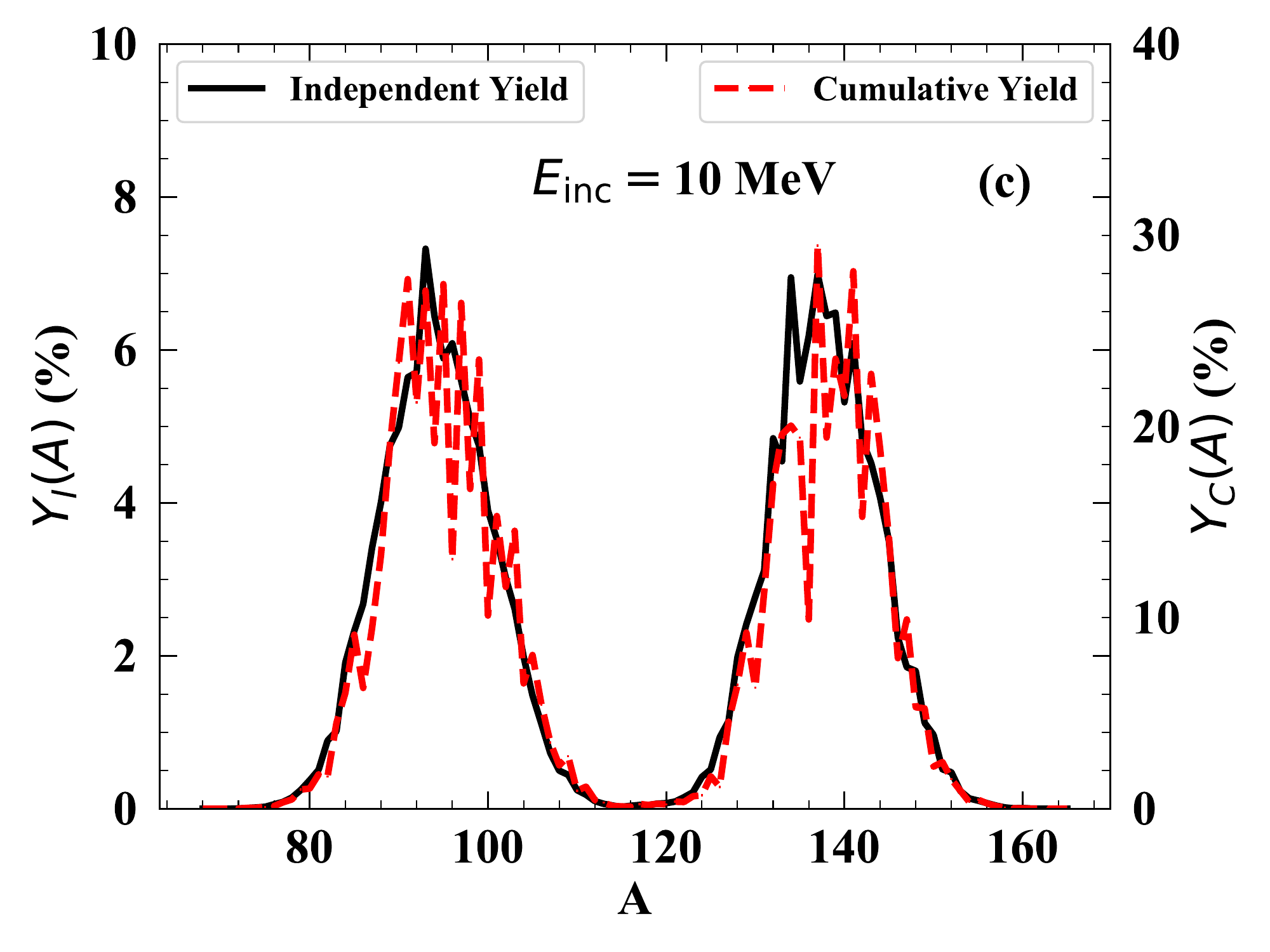} \\
\includegraphics[width=0.33\textwidth]{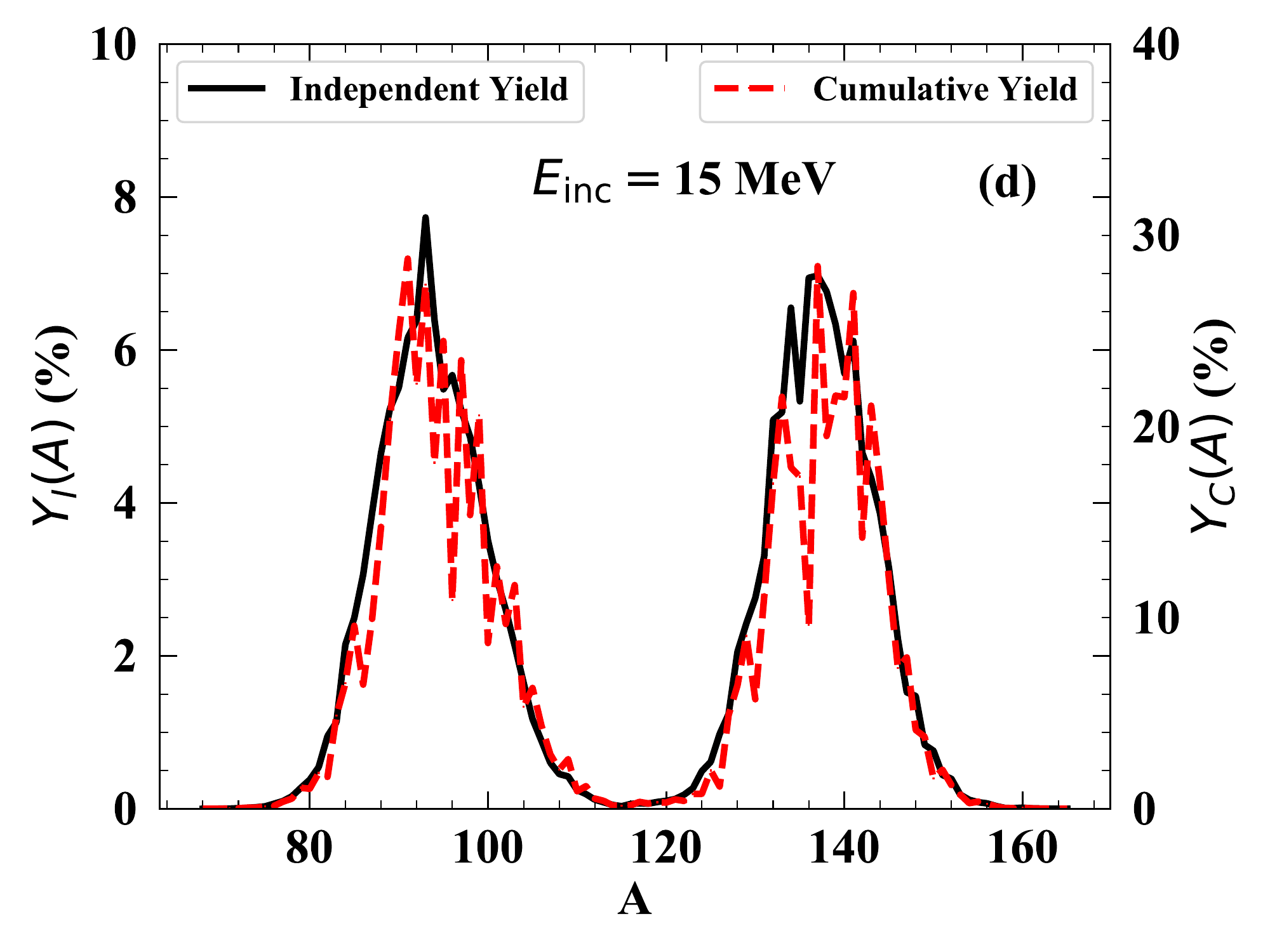} & \includegraphics[width=0.33\textwidth]{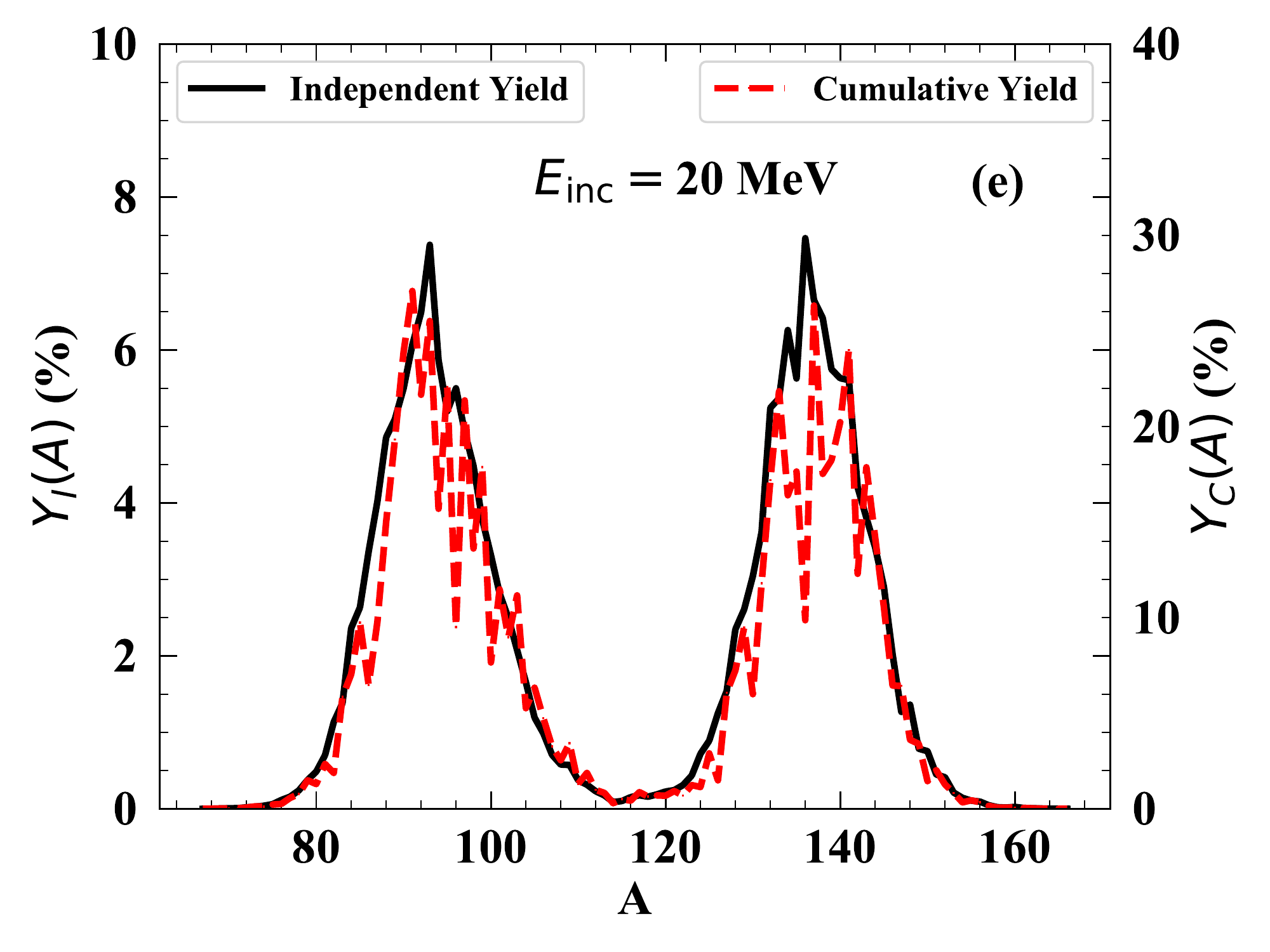} & \\
\end{tabular}
\caption{Independent fission mass yields (black solid) and cumulative fission mass yields (red dashed) for $^{235}$U(n,f) for incident energies of (a)~thermal, (b)~5 MeV, (c)~10 MeV, (d)~15 MeV, and (e)~20 MeV.  Note that the scale for the independent yields is on the left, and the scale for the cumulative yields is on the right.}
\label{fig:YIA}
\end{figure*}

In \beoh, we record the independent yields after the prompt neutrons are emitted.  Although we keep track of the meta-stable states, we can reconstruct the full independent yields, as a function of mass and charge, $Y_I(A,Z)$, or just as a function of mass, $Y_I(A)$.  $Y_I(A)$ calculated from \beoh{} with thermal-induced neutrons are shown in Fig. \ref{fig:YIA}(a), black solid line.  The evolution of the independent mass yields as the incident neutron energy increases is shown in the remaining panels of Fig. \ref{fig:YIA}:  (b) $E_\mathrm{inc}=$ 5 MeV, (c) $E_\mathrm{inc}=$ 10 MeV, (d) $E_\mathrm{inc}=$ 15 MeV, and (e) $E_\mathrm{inc}=$ 20 MeV.  Then, following the procedure in Sec. \ref{sec:delayedCalc}, we calculate the cumulative fission yields using the ENDF/B-VIII.0~\cite{ENDFB8} decay data library.  The cumulative mass yields are calculated, for the same incident energies, and compared to the independent mass yields in the corresponding panels of Fig. \ref{fig:YIA}.  

\begin{figure*}
\centering
\begin{tabular}{ccc}
\includegraphics[width=0.33\textwidth]{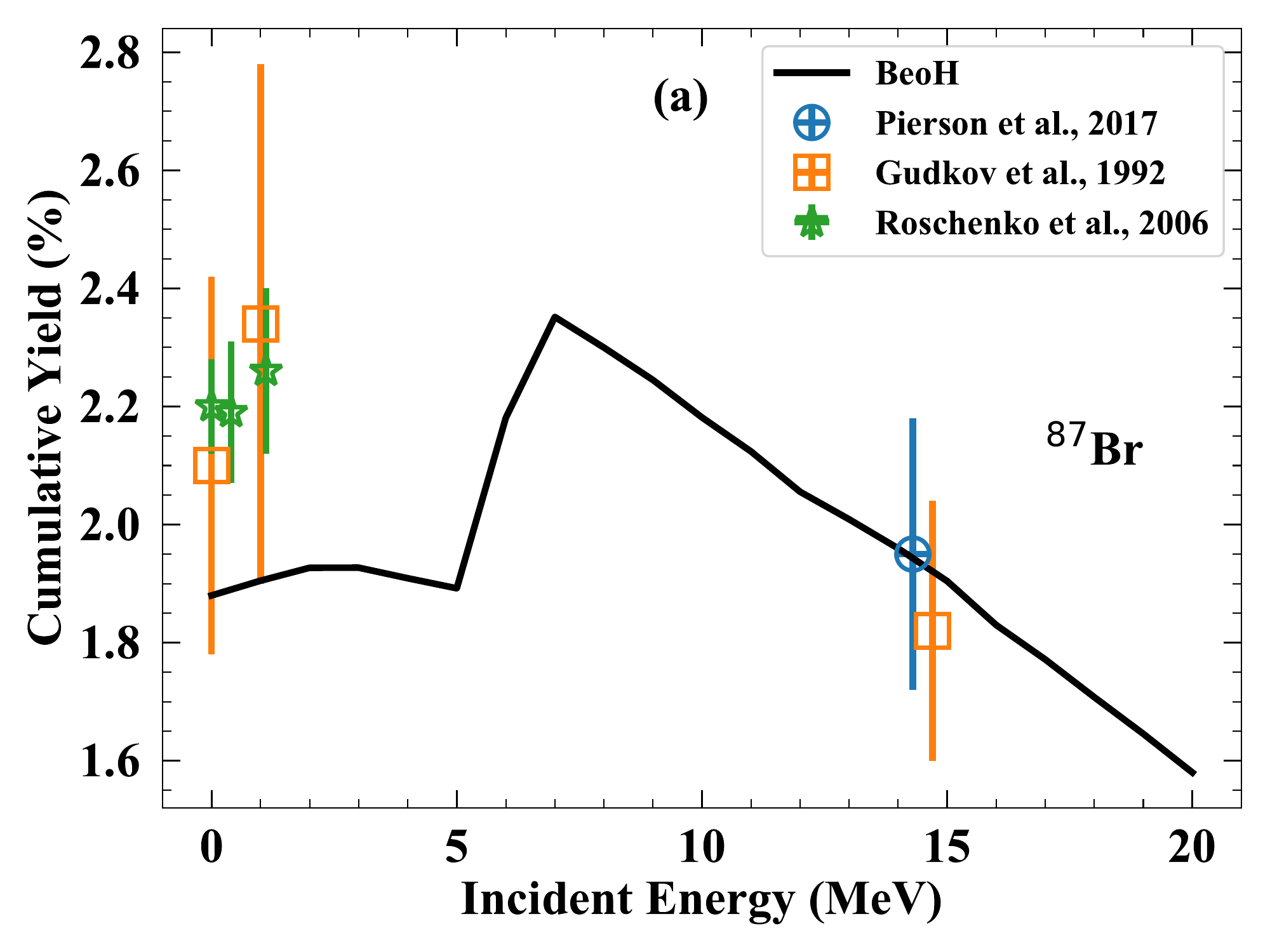} & \includegraphics[width=0.33\textwidth]{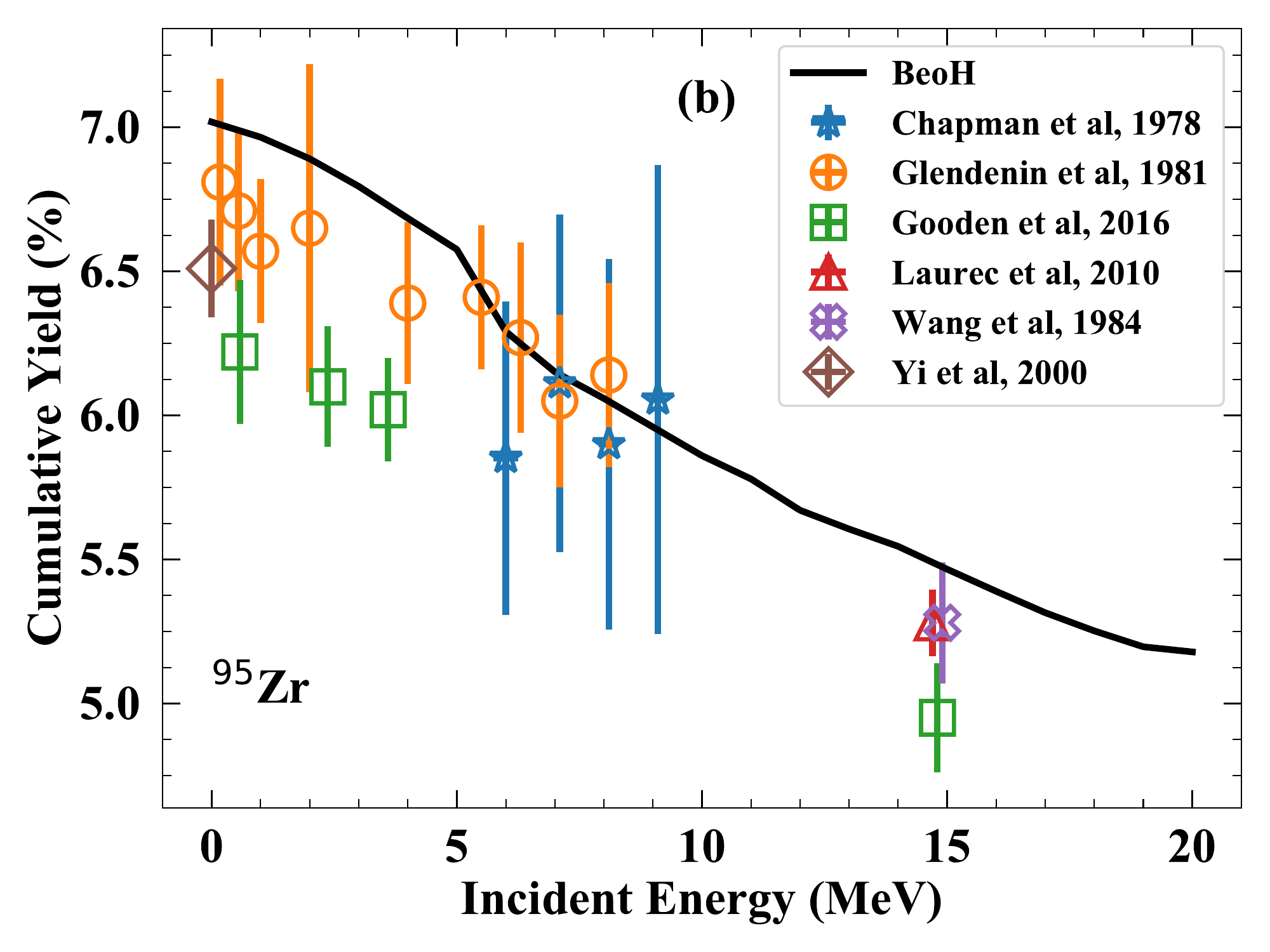} & \includegraphics[width=0.33\textwidth]{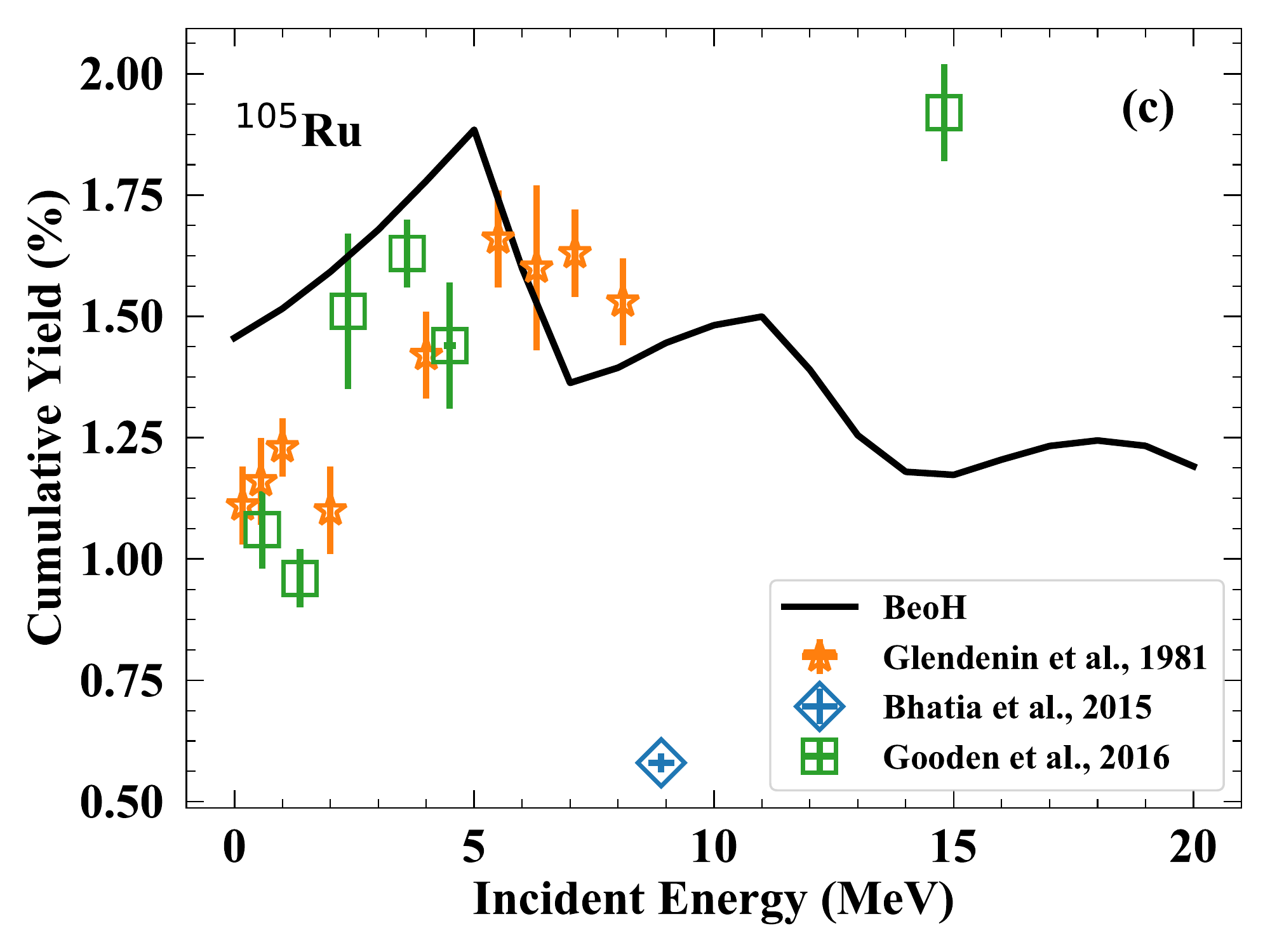} \\ 
\includegraphics[width=0.33\textwidth]{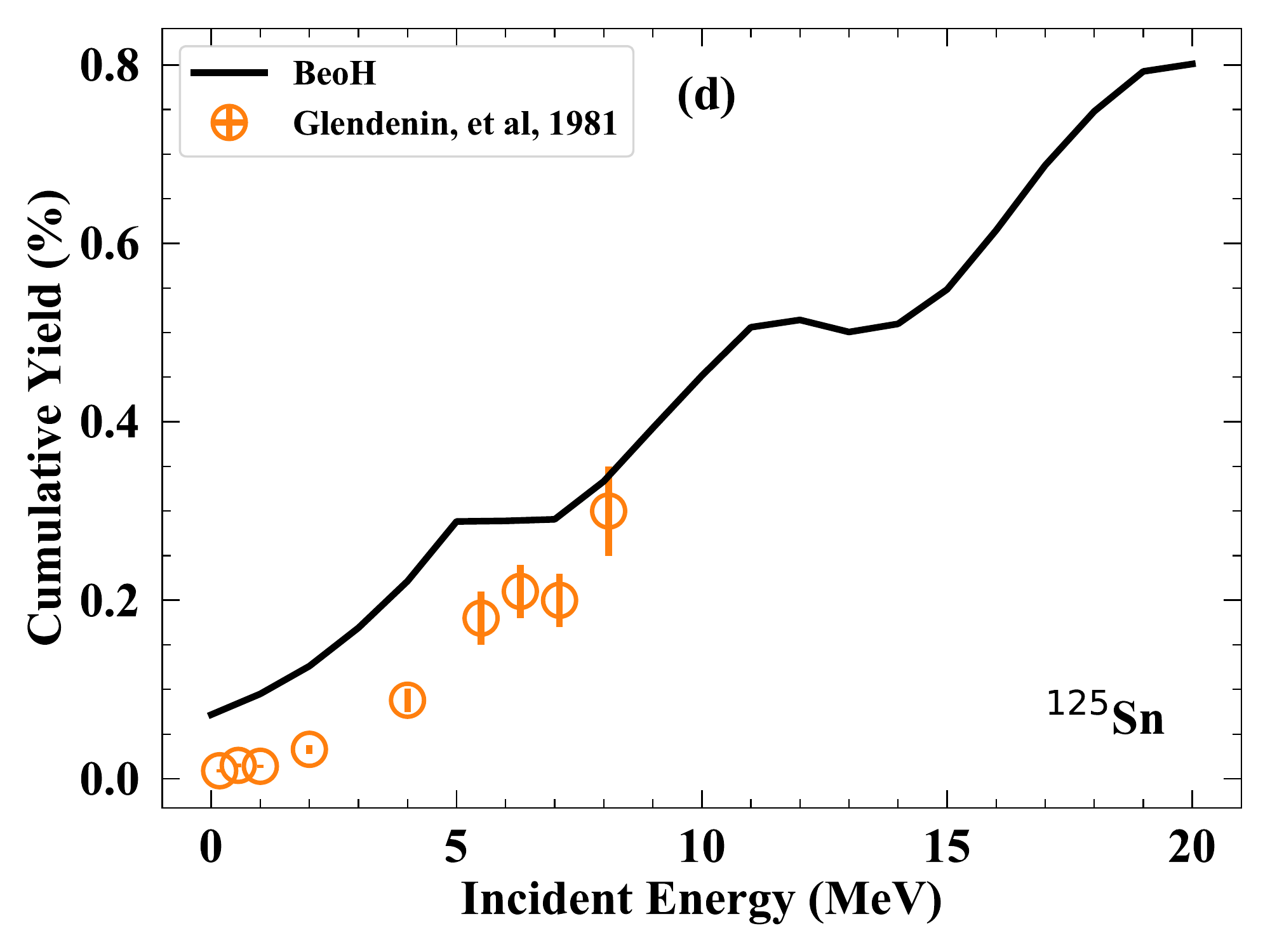} & \includegraphics[width=0.33\textwidth]{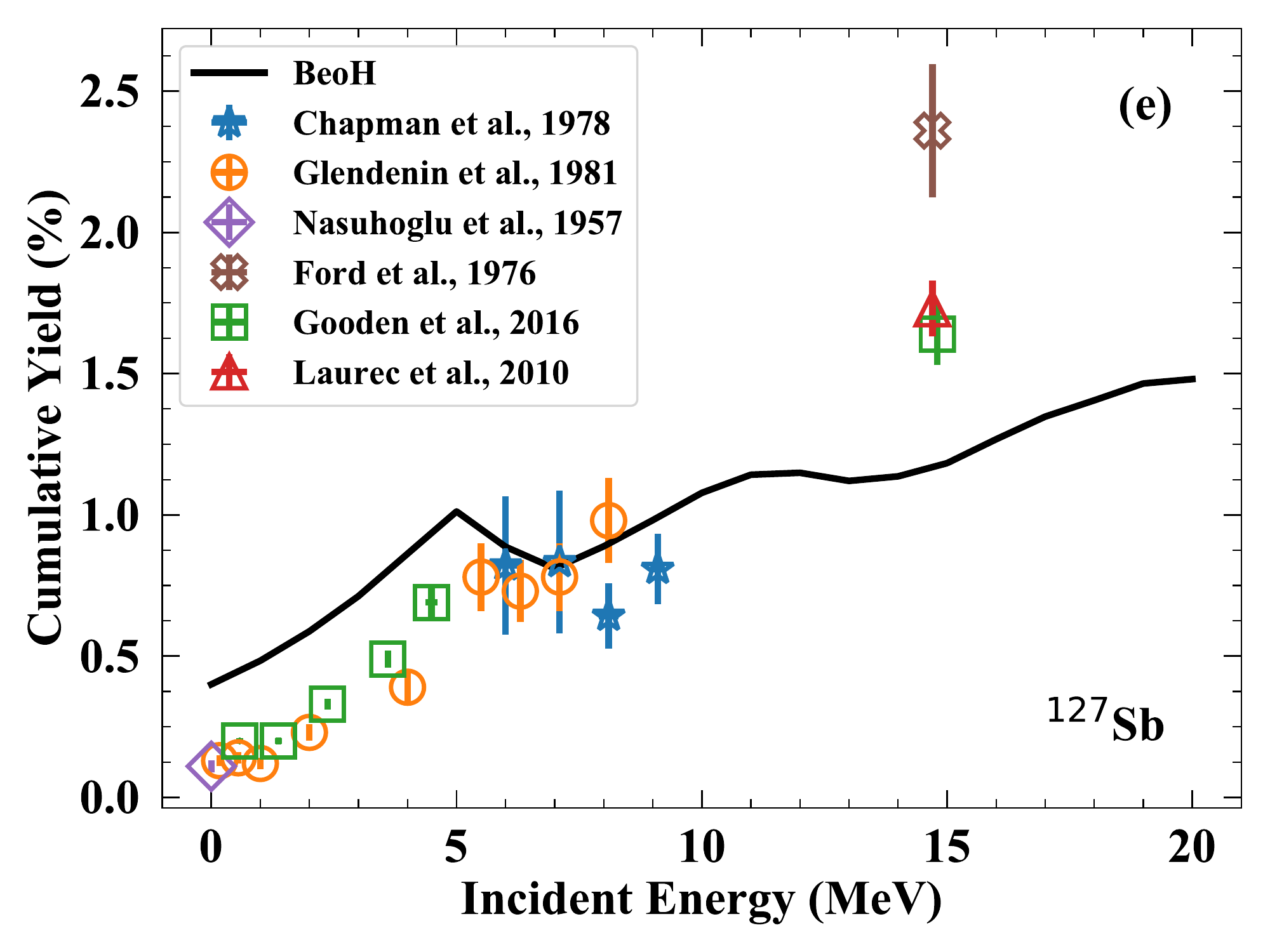} & \includegraphics[width=0.33\textwidth]{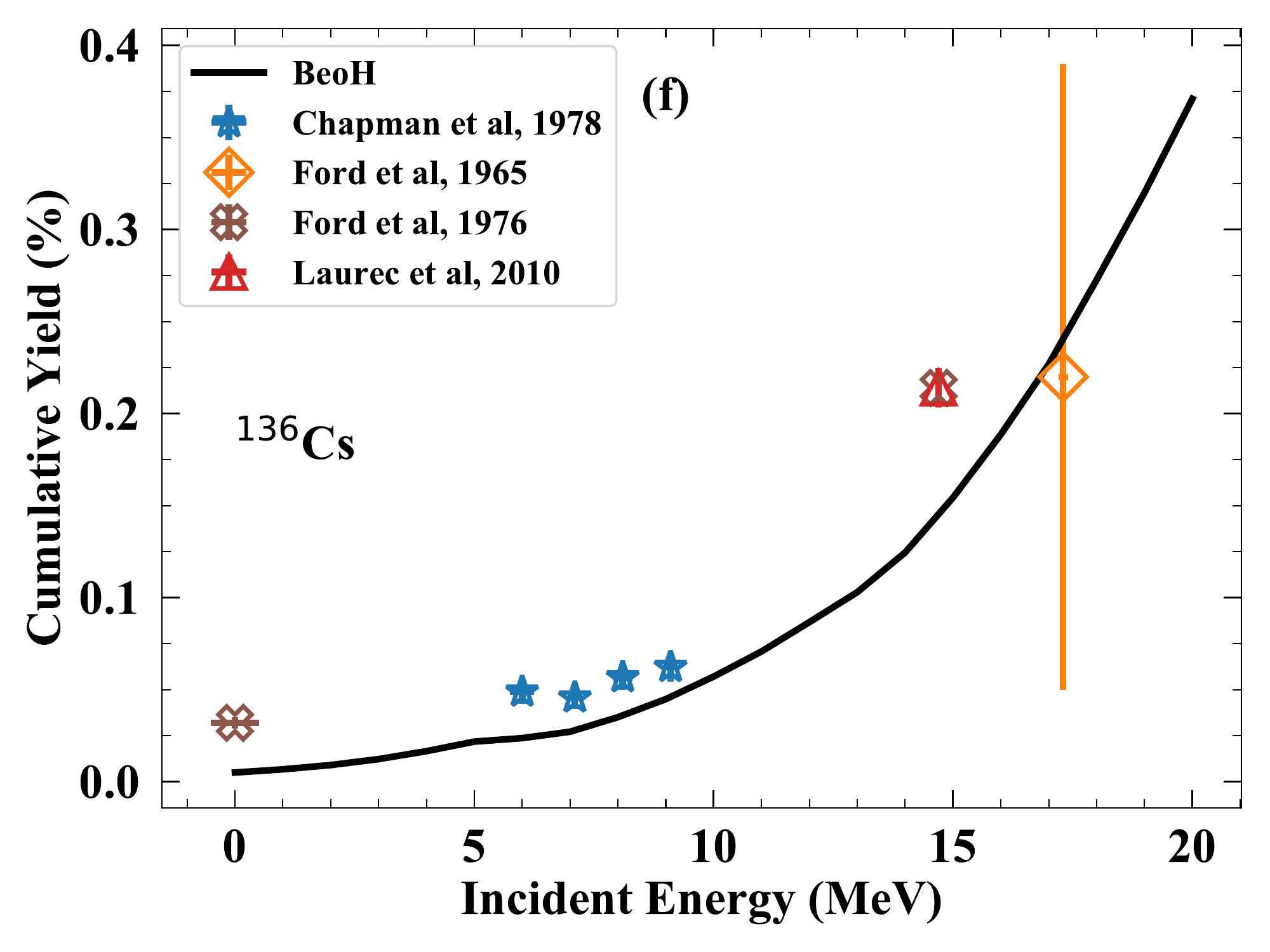} \\
\end{tabular}
\caption{Cumulative fission yields for select isotopes compared to available experimental data \cite{Pierson2017,Gudkov1992,Roshchenko2006,Chapman1978,Glendenin1981,Gooden2016,Laurec2010,Xiuzhi1984,Yi2000,Bhatia2015,Nasuhoglu1957,Ford1976,Ford1965}, (a)~$^{87}$Br, (b)~$^{95}$Zr, (c)~$^{105}$Ru, (d)~$^{125}$Sn, (e)~$^{127}$Sb, and (f)~$^{135}$Cs.}
\label{fig:CFY}
\end{figure*}

Further, in Fig. \ref{fig:CFY}, we show examples of select cumulative fission yields, summed over all isomeric states, as a function of the incident neutron energy.  We have chosen to show isotopes for which there is ample experimental data across a range of incident neutron energies.  For these isotopes, the \beoh{} calculations follow the shape of the experimental data as a function of incident energy, even though these data were not included in determining the parametrization for the pre-neutron yield distribution.  Even for isotopes where the magnitude of the \beoh{} calculations does not reproduce the experimental data exactly, the energy-dependent trends show the fidelity of the model.  In addition, there is still flexibility within the parameter space to further optimize the models needed for the pre- and post-scission calculations, particularly where there are few experimental measurements or measurements cannot be made directly.  

\begin{figure}
\centering
\includegraphics[width=0.5\textwidth]{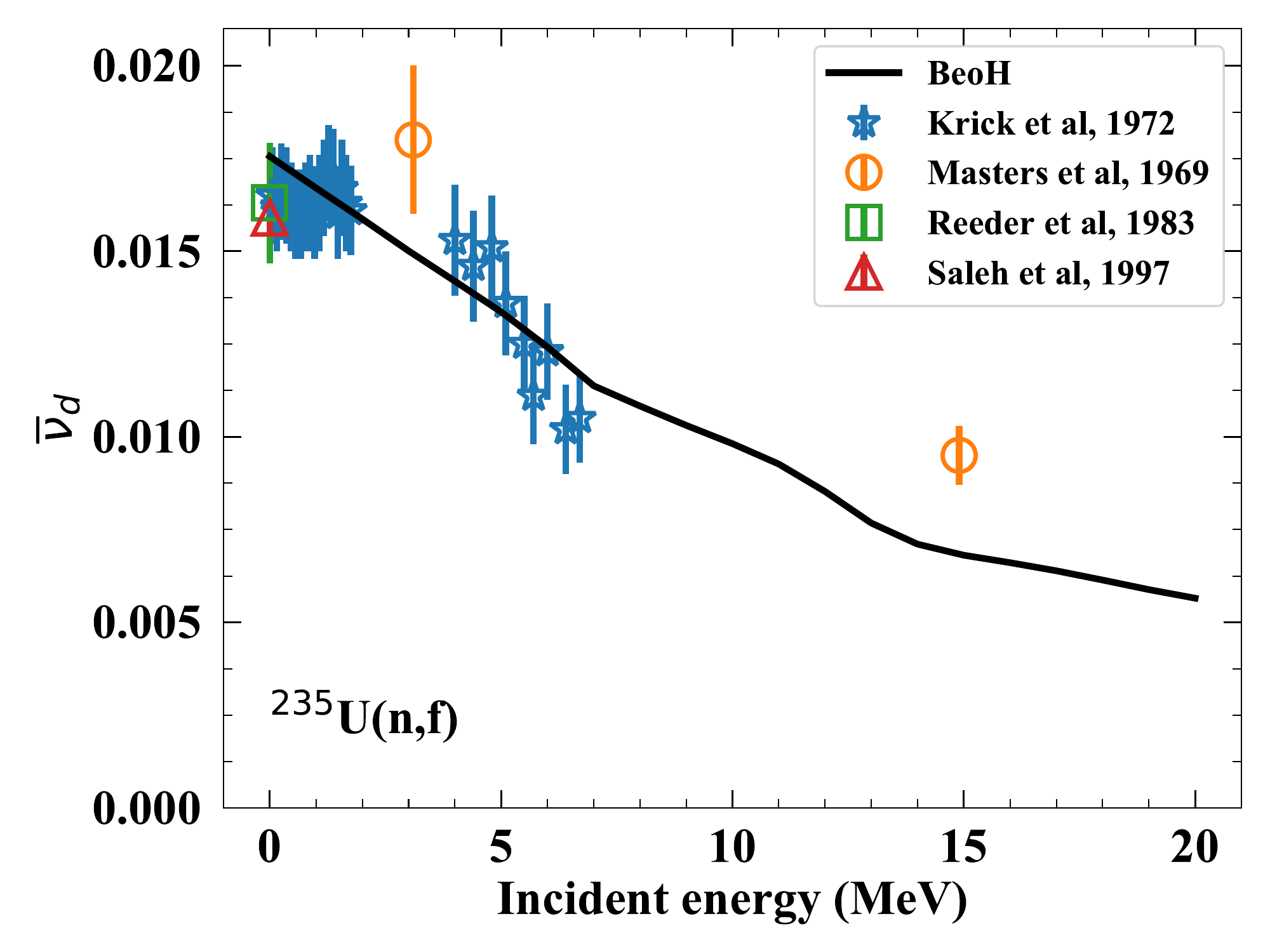}
\caption{Average number of delayed neutrons calculated by \beoh{} (black solid) compared to experimental data \cite{Krick1972,Masters1969,Reeder1983,Saleh1997}.}
\label{fig:nubarD}
\end{figure}

Finally, we calculate the average number of delayed neutrons emitted, shown in Fig. \ref{fig:nubarD}, compared to experimental data.  Above incident energies of about 2 MeV, \beoh{} reproduces the magnitude and trend of the experimental data, however, we do not see a flattening of $\overline{\nu}_d$ below this energy.  Preliminary studies show that a slope change in $\langle \mathrm{TKE} \rangle (E_\mathrm{inc})$, as seen in the data of Duke, et al.,~\cite{Duke2015} can lead to a flattening of both $\overline{\nu}_p$ and $\overline{\nu}_d$ for incident energies before the slope change.  In addition, an inclusion of energy-dependent changes to the even-odd factors in the Wahl distribution also can flatten out the delayed neutron yield.  More investigations into these model updates are currently ongoing, with insights from microscopic and macroscopic-microscopic calculations of fission fragment mass and charge yields \cite{Verriere2019}.

\section{Conclusions}
\label{sec:conclusions}


In summary, we have extended the deterministic Hauser-Feshbach fission fragment decay model within the code \beoh{} to calculate prompt and delayed particle emission from fission fragments.  Using pre-scission inputs, such as multi-chance fission probabilities, average energy causing fission, and average pre-fission neutron energy, and post-scission inputs, such as the mass, charge, total kinetic energy, spin, and parity yields, fission fragment initial conditions are determined and then decayed through the Hauser-Feshbach statistical model, emitting prompt neutrons and $\gamma$ rays.  Prompt quantities, such as average multiplicities and energy spectra, are calculated from the decay of each fission fragment.  The independent fission product yields are determined, including any isomeric states.  Finally, cumulative fission yields are calculated through a time-independent calculation using decay data libraries, providing calculations of the $\beta$-delayed neutron yields.  Here, we show the first calculations for $^{235}$U(n,f) up to 20-MeV incident-neutron energy, which show a good agreement with energy-dependent prompt and delayed observables.

As the goal of this work is to show the developments in the \hfd{} model beyond first-chance fission, there is still work to be done in further optimizing the input parameters, not only at thermal neutron energies but as the incident energy increases as well - including taking into account the change in the fissioning compound nucleus as the multi-chance fission channels open.  In addition, while preliminary input are available for other major actinides, primarily $^{238}$U and $^{239}$Pu, these need to be tested against delayed observables and further updated.  Both of these tasks are currently underway.  
 

\begin{acknowledgements}
The authors would like to thank T.~Bredeweg, M.~Gooden, M.~B.~Chadwick, and A.~Tonchev for useful discussions.  This work was performed under the auspice of the U.S. Department of Energy by Los Alamos National Laboratory under Contract 89233218CNA000001 and was supported by the Office of Defense Nuclear Nonproliferation Research \& Development (DNN R\&D), National Nuclear Security Administration, U.S. Department of Energy.  
\end{acknowledgements}


\appendix

\section{Multi-chance fission details}
\label{app:MCFdetails}


Keeping track of all possible pairs of fission fragments for both the first-chance and multi-chance fission is rather technical, and therefore, we have not included the details in the main text.  The possible masses run from $A_\mathrm{min}=50$ to $A_\mathrm{max}=A_c - A_\mathrm{min}$.  For each possible mass, the most likely charge, $Z_p$, is determined from the Wahl systematics \cite{Wahl2002}, and the distribution $P(Z|A)$ is calculated for charges in the range $Z_p \pm 5$.  For first-chance fission, the mass and charge yield is symmetric with respect to $A_c/2$,
\begin{equation}
Y(Z_l,A_l) = Y(Z_c-Z_l,A_c-A_l) = Y(Z_h,A_h).
\end{equation} 
All possible pairs are generated where
\begin{eqnarray}
Y(k) = Y(Z_l,A_l,Z_h,A_h) & = & Y(A_l)P(Z_l|A_l), \\
& = & Y(A_h)P(Z_h|A_h) \nonumber
\end{eqnarray}
and $k$ is the pair index.  When $Y(k)$ is below the desired yield cut-off, that pair is discarded.

Above first-chance fission, it would be most straight-forward to run \beoh{} for each fission chance then weight the resulting calculations by the fission probabilities.  However, this procedure increases the computational time $m$-fold.  Instead, the calculation steps are rearranged slightly.  All fission pairs for each of the fissioning compounds are first calculated, $Y(m,k)$, where $A_l+A_h=A_c-m+1$.  For first-chance ($m=1$), this is the same as the above calculation, however, for second-chance fission and above ($m\ge 2$), the mass and charge yields are calculated at the equivalent incident energy, $E_\mathrm{eq}(m)$, from Eq. (\ref{eqn:Eeq}), along with the total kinetic energy distribution.  $Y(1,k)$ is then expanded to include all possible fission pairs that occur in all $Y(m,k)$ calculations; if this pair appears only for $m>1$, its pair yield is set to zero.  

For each $(Z_l,A_l)$-$(Z_h,A_h)$ pair in $Y(1,k)$, each other $Y(m,k^\prime)$ is scanned to see if $(Z_l,A_l)$ or $(Z_h,A_h)$ appears, and if that fission fragment appears at a different fission chance, its pair yield is added to either the light or heavy fragment.  For example, if a pair $(Z_l,A_l)$-$(Z_h,A_h)$ appear in $Y(1,k)$, and the pair $(Z_l,A_l)$-$(Z_h,A_h-1)$ appears in $Y(2,k^\prime)$, the light fragment yield would be $P_f(1)Y(1,k)+P_f(2)Y(2,k)$ while the heavy fragment yield would be $P_f(1)Y(1,k)$ because the heavy fragment is not the same in the first-chance and second-chance calculations due to the change in the pre-scission compound nucleus.  The fragment yields then become,
\begin{eqnarray}
Y_l(k) & = &  \sum _{m,k^\prime} Y(m,k^\prime) P_f(m) \delta (A_l,A_l^\prime) \delta (Z_l,Z_l ^\prime) \\
& + &  \sum _{m,k^\prime} Y(m,k^\prime) P_f(m) \delta (A_l,A_h^\prime) \delta (Z_l,Z_h ^\prime) \nonumber
\label{eqn:Ylk}
\end{eqnarray}
for the light fragment and 
\begin{eqnarray}
Y_h(k) & = &  \sum _{m,k^\prime} Y(m,k^\prime) P_f(m) \delta (A_h,A_h^\prime) \delta (Z_h,Z_h ^\prime) \\
& + &  \sum _{m,k^\prime} Y(m,k^\prime) P_f(m) \delta (A_h,A_l^\prime) \delta (Z_h,Z_l ^\prime) \nonumber
\label{eqn:Yhk}
\end{eqnarray}
for the heavy fragment.

Each $m^\mathrm{th}$-chance fragment pair has had its TKE and TXE calculated already, and the average excitation energy, $E_{l,h}(m)$, and width, $\delta_{l,h}(m)$, are calculated independently.  The statistical Hauser-Feshbach decay from a given excitation energy $E_x$ is independent of the decay from other excitation energies $E_x ^\prime$.  Then because the statistical Hauser-Feshbach decay from excitation energies $E_x$ and $E_x^\prime$ are independent from one another, for the multi-chance fission case, the initial population distribution can be constructed a superposition of multi-chance component $G(m,E_x)$ weighted by the fission probability $P_f(m)$.  The constructed energy distribution is then
\begin{eqnarray}
G_l(E_x) & = & \sum _{m,k^\prime} P_f(m) G_l(m,E_x) \delta (A_l,A_l^\prime) \delta (Z_l,Z_l ^\prime) \\
& + & \sum _{m,k^\prime} P_f(m) G_l(m,E_x) \delta (A_l,A_h^\prime) \delta (Z_l,Z_h ^\prime) \nonumber
\end{eqnarray}
and 
\begin{eqnarray}
G_h(E_x) & = &  \sum _{m,k^\prime} P_f(m) G_h(m,E_x) \delta (A_h,A_h^\prime) \delta (Z_h,Z_h ^\prime) \\
& + &  \sum _{m,k^\prime} P_f(m) G_h(m,E_x) \delta (A_h,A_l^\prime) \delta (Z_h,Z_l ^\prime) \nonumber
\end{eqnarray}
similar to Eqs. (\ref{eqn:Ylk}) and (\ref{eqn:Yhk}), with 
\begin{equation}
G_{l,h}(m,E_x) = \frac{1}{\sqrt{2\pi}\delta _{l,h}(m)} \exp \left [ - \frac{(E_x - E_{l,h}(m))^2}{2\delta_{l,h}^2(m)} \right ].
\end{equation}
The spin distribution is calculated in the same fashion.  This technique allows us to reduce the $m$-fold computation ($m$-times Hauser-Feshbach decay calculations) into just one.

\section{Optimized parameter values}
\label{app:parmValues}

In Table \ref{tab:parameters}, we show the pre-neutron yield parameters that were used in the calculations presented here.  Each column is labeled with the compound nucleus for first- ($^{236}$U), second- ($^{235}$U), and third- ($^{234}$U) chance fission.  When fourth-chance fission is energetically allowable, the third-chance parametrization is used.

\begin{table}
\centering
\begin{tabular}{c|c|c|c}
& $^{236}$U & $^{235}$U & $^{234}$U \\ \hline \hline
$w_1^a$ & -6.856 & -31.199 & 30.161 \\
$w_1^b$ & 6.082 & 30.000 & 19.558 \\
$\sigma _1 ^a$ & 15.790 & 16.600 & 15.680 \\
$\sigma _1 ^b$ & -0.280 & -0.020 & -0.030 \\
$\mu _1 ^a$ & 3.029 & 3.751 & 3.696 \\
$\mu _1 ^b$ & 0.000 & 0.082 & 0.000 \\
$w_2^a$ &  -6.864 & -30.578 & -29.524 \\
$w_2^b$ & -6.1438 & -29.3655 & -19.3867 \\
$\sigma _2 ^a$ & 22.970 & 24.120 & 23.310 \\
$\sigma _2 ^b$ & -0.270 & -0.040 & -0.050 \\
$\mu _2 ^a$ & 4.694 & 5.037 & 5.107 \\
$\mu _2 ^b$ & 0.185 & 0.109 & 0.025 \\
$\sigma _0 ^a$ & 9.885 & 10.011 & 10.302 \\
$\sigma _0 ^b$ & 0.032 & 0.110 & 0.000\\
$a$ & 170.600 & 170.200 & 168.800 \\
$E_0$ & 0.000 & 0.000 & 0.000 \\
$b$ & 0.000 & 0.000 & 0.000 \\
$d$ & -0.180 & -0.200 & -0.220 \\
$p_0$ & 324.690 & 324.690 & 324.690 \\
$p_1$ & 1.101 & 1.101 & 1.101 \\
$p_2$ & 0.1768 & 0.1768 & 0.1768 \\
$p_3$ & 40.509 & 40.509 & 40.509 \\
$s_0$ & 5.520 & 5.520 & 5.520 \\
$s_1$ & 5.750 & 5.750 & 5.750 \\
$s_2$ & 5512 & 5512 & 5512 \\
$f$ & 2.756 & 2.756 & 2.756 \\
$f_Z$ & 1.073 & 1.073 & 1.073 \\
$f_N$ & 0.998 & 0.998 & 0.998 \\
\end{tabular}
\caption{Fitted parameters for pre-neutron emission yields used in this work, for the $^{236}$U, $^{235}$U, and $^{234}$U compound nuclei.  For the highest incident energies, where fourth-chance fission is energetically allowed, the $^{234}$U compound parameters are used.  Energies ($a$ and $E_0$ are in MeV.}
\label{tab:parameters}
\end{table}

It is worth noting that the parameters for mass yield, $Y(A|E_\mathrm{inc})$, can be degenerate (especially for the $w_i$ parameters) and often hits the boundaries of the defined parameter space when fitting to experimentally measured mass yields.  Further studies should be performed to understand if these degeneracies have an effect on the resulting prompt and delayed observables.


\bibliography{FPY}

\end{document}